\documentclass[a4paper,11pt]{article}
\pdfoutput=1 

\usepackage{jcappub} 

\usepackage[T1]{fontenc} 

\title{\boldmath Analytical investigation of pre-inflationary effects in the primordial power spectrum: From General Relativity to hybrid Loop Quantum Cosmology}

\author[a,1]{Beatriz Elizaga Navascu\'es,\note{Corresponding author.}}
\author[b]{Guillermo A. Mena Marug\'an}

\affiliation[a]{JSPS International Research Fellow, Department of Physics, Waseda University,\\3-4-1 Okubo, Shinjuku-ku, 169-8555 Tokyo, Japan}
\affiliation[b]{Instituto de Estructura de la Materia, IEM-CSIC,\\Serrano 121, 28006 Madrid, Spain}

\emailAdd{w.iac20060@kurenai.waseda.jp}
\emailAdd{mena@iem.cfmac.csic.es}

\abstract{This work is an analytical study of the main differences between classical inflationary effects of a fast-roll regime and imprints of hybrid Loop Quantum Cosmology (LQC) on the primordial power spectrum. Effective LQC solutions of phenomenological interest typically contain a classical period of kinetic dominance prior to slow-roll inflation, with consequences on the power spectrum that here we try and tell apart from those of phenomena occuring in the vicinity of a bounce in hybrid LQC. For this comparison, we use the same criterion for the choice of a vacuum state of the primordial perturbations, both in the context of relativistic cosmology and on the LQC cosmological background. As a first approximation to the problem, our study ignores the influence of a quasi-de Sitter evolution (rather than de Sitter) in the slow-roll inflationary regime, and of a typically short transition between the kinetically dominated era and the slow-roll period. Our results show that, while the two studied types of effects can lead to a drastic power suppression in the infrared sector of the spectrum, the mode scale at which this happens is much larger for LQC than for classical relativistic cosmology, the corresponding scales being related to the spacetime curvature at the bounce and at the onset of inflation, respectively.}


\begin{document}
\maketitle
\flushbottom

\section{Introduction}
\label{sec:intro}

The Cosmic Microwave Background (CMB) is a useful observational window to the earliest stages of the Universe. Its angular power spectrum of temperature (and polarization) anisotropies can be explained remarkably well starting from a primordial slow-roll inflationary epoch, using standard cosmological perturbation theory in General Relativity (GR) \cite{structures,mukhanov1}. Despite this success, both WMAP \cite{wmap} and, more recently, the Planck mission \cite{planck,planck-inf} have reported an anomalous lack of power for the lowest multipoles of the observable spectrum, namely for the correlations between points in the sky separated by the largest angles. Even though this sector of the power spectrum is strongly affected by cosmic variance, this anomalous behavior may open the possibility for new physics, beyond the otherwise successful inflationary paradigm.

During last years, a considerable number of works have proposed a primordial origin for this anomalous lack of power in the CMB spectrum, without giving up on a relativistic description of the early Universe (according to GR) nor on its spatial flatness \cite{CPKL,CCL,WNg,DVS1,DVS2,BDVS,JCGSS,SGC,NFS,DKPS,KKO,PL,HHHL}. Instead, these investigations often relax the hypothesis that no information on the pre-inflationary Universe survives the slow-roll epoch (this point of view has been reinforced recently  e.g. through the study of the path integral formulation of quantum cosmology \cite{Turok}). Such investigations often consider scenarios in which the slow-roll period is just long enough to satisfy the observational constraints on the necessary e-folds \cite{RS,Ramirez,SA}. In many cases, this situation is attained by considering inflationary models with multiple, different stages. Among these, perhaps the simplest ones are those that, driven by a single scalar field, contain what has been called a phase of fast-roll inflation (see e.g. refs. \cite{CPKL,DVS1,DVS2,HHHL}). They describe a cosmological evolution with a (pre-inflationary) epoch in which the kinetic energy of the scalar field dominates over its potential energy, followed by a short period of slow-roll inflation. With appropriate choices for the vacuum state of the cosmological perturbations, the primordial power spectrum resulting from these models displays a strong suppression, which can be interpreted as if there existed an effective infrared cutoff.

Another possibility to address the presence of anomalies in the CMB arises naturally in the context of quantum cosmology. As one approaches the Big-Bang singularity and the spacetime curvature grows unboundedly large, our conventional knowledge of physics indicates that the classical relativistic description eventually breaks down, and we expect that it should be replaced with an underlying quantum theory. The modifications coming from quantum geometry effects in the pre-inflationary evolution of the Universe could naturally involve phenomena leading e.g. to power suppression in the CMB. In this framework, an interesting scenario is provided by Loop Quantum Cosmology (LQC) \cite{bojo,abl,ashparam}. Apart from developing a non-perturbative quantization of cosmological spacetimes, LQC provides a mechanism for the resolution of the Big-Bang singularity into a bounce of quantum origin \cite{APS,mmo}. Certain Gaussian states of the theory effectively describe cosmological spacetimes that, evolving as in standard GR cosmology for low densities, depart from the classical trajectories when the energy density approaches the Planck scale and then avoid the singularity \cite{ashparam,taveras,jorma}. 

In the presence of an inflaton field, there is a class of especially appealing solutions in the aforementioned effective LQC regime. These are solutions with slow-roll inflation for which the amount of expansion from the bounce to the end of inflation is such that the perturbations with largest observable wavelengths (and only them) felt the quantum curvature effects near the bounce. Remarkably, these solutions in effective LQC display, shortly after the bounce, a classical period that mimmicks fast-roll inflation \cite{Ivan,hybr-pred,AGvacio2}. Namely, a pre-inflationary regime of kinetic dominance that eventually leads to a period of slow-roll inflation with just enough e-folds. Focusing on this type of solutions, a number of studies have predicted power suppression in the primordial spectrum that alleviaties the CMB anomalies \cite{hybr-pred,AGvacio2,no,AshNe}. In order to guarantee the potential falsifiability of these models, it then becomes very important to be able to distinguish whether such power suppression is due to genuine LQC effects on the evolution of the perturbations or to fast-roll effects (or to a combination of both). Unfortunately, all the studies of power suppression in LQC have so far relied on numerical computations, which fail to tell apart in detail the actual processes (classical or quantum) underlying their predictions on the modifications of the spectrum.

The aim of this work is to develop an analytical understanding of how the different types of pre-inflationary effects that coexist in the LQC evolution of scalar and tensor perturbations become reflected in the primordial power spectrum. More specifically, in this article we will concentrate our discussion exclusively on a concrete proposal for the description of cosmological perturbations in LQC: the hybrid approach \cite{hyb-pert2,hyb-pert3,hyb-pert4,hybr-ten}. This is one of the proposals from which suppression for large wavelengths of the primordial power spectrum has been reported. An important reason to focus our analysis on hybrid LQC is that it provides a manageable and well-motivated framework to try and disentangle the ultimate origin of the different phenomena that affect the power spectra. Nevertheless, we believe that the techniques and intuition gained with this analysis can be extended to other theoretical frameworks, facilitating a better understanding of the observational consequences of pure relativistic effects and trans-Planckian physics in the very early Universe (even beyond the context of LQC). The dynamical equations for the perturbations in hybrid LQC are derived from a quantization of the full cosmological system that combines loop quantization techniques for the spatially homogeneous degrees of freedom (the cosmological background) with a Fock quantization for the perturbations. These dynamical equations have the same hyperbolic structure as in leading-order cosmological perturbation theory for GR, but display a time-dependent mass term that differs from the classical one in the vicinity of the bounce \cite{mass}. We will analytically study these equations for background solutions of effective LQC in situations of phenomenological interest, on the one hand, and for their counterpart given by fast-roll inflationary solutions in GR cosmology, on the other hand. Then, using the same criterion for the choice of vacuum state in both cases (a choice that is motivated by physical considerations), we will compare the resulting power spectra and study the differences, which can be rooted in the existence of two distinct scales in the considered situations, namely, the spacetime curvature at the bounce, with strong LQC effects, and the curvature at the onset of slow-roll inflation, which is ruled by GR cosmology.

In order to carry out this analytical study, we will need to approximate in a convenient form the time-dependent mass that appears in the dynamical equations of the perturbations. For that purpose, we will model both the GR cosmological solution with fast-roll inflation, and the part of the LQC background with a similar classical behavior, as a cosmological spacetime minimally coupled to a massless scalar field that instantaneously transitions to a de Sitter spacetime. In this respect, as a first approximation to the problem, we will ignore any effect produced during the passage from kinetic dominance to slow-roll inflation. Furthermore, we will also neglect the departures from an exact de Sitter expansion during slow-roll inflation, which depend on the specific inflaton potential that one is considering. As a consequence, our study will not provide any reliable detail about slow-roll observables such as the spectral index or the tensor-to-scalar ratio. On the other hand, in the vicinity of the LQC bounce, we will approximate the time-dependent mass as certain P\"oschl-Teller potential, in a way that improves similar approximations made in the literature of (hybrid) LQC \cite{wang} in what respects the time interval of applicability. Although limiting the extent and accuracy of our physical investigation, the simplicity of these approximations will actually allow us to deduce and clearly explain the main differences between the primordial power spectra obtained with pure GR cosmology and with effective hybrid LQC.

The structure of the rest of this paper is the following. In section \ref{sec:ii} we approximate and discuss analytically the propagation of cosmological perturbations on a single-field cosmology with fast-roll inflation within GR. Then, we define the primordial power spectrum and comment on some of its properties. The general strategy that we follow in this section is the same as in ref. \cite{CPKL}, but maintaining the generality in the choice of vacuum state for the perturbations. In section \ref{sec:iii} we revisit the properties of the effective LQC solutions of phenomenological interest, and make a P\"oschl-Teller approximation in the vicinity of the bounce to approximate the dynamics of the cosmological perturbations in the hybrid approach. We then explain in which sense this approximation improves similar ones in the literature \cite{wang}. In section \ref{sec:iv} we set a common criterion for the choice of vacuum state, with original motivation coming from the hybrid quantization process itself and the demand of non-oscillatory properties on the power spectrum. We use this criterion to select a state for the cosmological perturbations, both in the GR cosmological model and in hybrid LQC. We also analytically compute and compare the resulting power spectra. Finally, in section \ref{sec:v} we discuss our results and their prospective applications. Throughout the paper we work in Planck units, setting $G=c=\hbar=1$.

\section{Perturbations in de Sitter inflation matching a scalar field pre-inflationary phase}
\label{sec:ii}

Let us set up the theoretical model with which we want to approximate a classical, single-field driven, inflationary universe preceeded by a classical epoch of deccelerated expansion, in the presence of inhomogeneous perturbations. This cosmological evolution takes place in an interval $(\eta_0,\eta_{\mathrm{end}}]$ of the conformal time $\eta$, where $\eta_{\mathrm{end}}$ is the instant when slow-roll inflation ends. For the moment, let $\eta_0$ be an unspecified choice of initial time. Ignoring any perturbative backreaction, the equations that rule the evolution of the considered background cosmology in conformal time are
\begin{equation}\label{backGR}
\left(\frac{a'}{a}\right)^{2}=\frac{8\pi }{3}a^2\rho, \qquad \frac{a''}{a}=\frac{4\pi }{3}a^2\rho-4\pi  a^2 P,
\end{equation}
where $a$ is the cosmological scale factor, the prime denotes the derivative with respect to the conformal time, and $\rho$ and $P$ are the respective energy density and pressure of the homogeneous inflaton field $\phi$ with potential $V(\phi)$, namely
\begin{equation}\label{rhop}
\rho=\frac{1}{2}\left(\frac{\phi'}{a}\right)^2+V(\phi) ,\qquad P=\rho-2 V(\phi).
\end{equation}
For concreteness, in this work we focus the discussion on the case of a quadratic potential $V(\phi)=m^2 \phi^2/2$, where $m$ is the inflaton mass. However, since the details of this potential during inflation will be completely ignored within our approximations, we do not expect our results to be extremely sensitive to this choice.

In the context of dominant-order perturbation theory, the equations of motion for the gauge-invariant perturbations on the considered background are 
\begin{align}\label{modexactgr}
v^{(r)\prime\prime}_{\vec{k}}+\left[k^2+s^{(r)}_{\text{GR}}\right]v^{(r)}_{\vec{k}}=0,\qquad k=|\vec{k}|,
\end{align}
where $v^{(r)}_{\vec{k}}$ generically denotes either the Fourier coefficient of the mode with wavevector $\vec{k}$ of the Mukhanov-Sasaki field \cite{mukhanov,sasaki,sasakikodama} (field for which we employ the notation $r=s$) or the respective coefficient of the tensor perturbations (case for which $r=t$). The function $s^{(r)}_{\text{GR}}$ plays the role of a time-dependent mass for the respective Mukhanov-Sasaki and tensor perturbations. Explicitly, it is well-known that \cite{langlois}
\begin{equation}\label{massgr}
s^{(s)}_{\text{GR}}=-\frac{z''}{z},\qquad s^{(t)}_{\text{GR}}=-\frac{a''}{a},
\end{equation}
where $z=a^2\phi'/a'$. Even for times when $\phi'=0$, the Mukhanov-Sasaki mass is non-singular and can be expressed in terms of the tensor one in a general relation of the form \cite{mass}
\begin{equation}
s^{(s)}_{\text{GR}}= s^{(t)}_{\text{GR}}+\mathcal{U}_{\text{GR}},
\end{equation}
where $\mathcal{U}_{\text{GR}}$ is a linear combination (with time-dependent coefficients) of the potential $V(\phi)$, its square, and its first and second derivatives with respect to $\phi$. We do not need to show the explicit expression of $\mathcal{U}_{\text{GR}}$ here, but only to notice that it is a very small term when $V(\phi)$ and its two first derivatives are negligible compared to the kinetic component of the energy density $\rho$.

\subsection{The approximate cosmological background}

Among the solutions of eq. \eqref{backGR} for the cosmological background in the period $(\eta_0,\eta_{\mathrm{end}}]$, we restrict our attention to those with a subinterval $(\eta_0,\eta_i)$ where the kinetic contribution to the energy density greatly dominates over the potential one. In this way, we focus the discussion on solutions that contain a period of fast-roll inflation. These are the only solutions from which one can expect non-trivial imprints of pre-inflationary classical physics in the largest observable angular scales of the CMB. On top of this behavior of early kinetic dominance, observability of these pre-inflationary effects requires that the slow-roll inflationary phase does not last long enough so as to shift the affected scales to regions of the spectrum that are not yet visible to us nowadays. Such inflationary processes often go by the name of just-enough or short-lived inflation \cite{Ramirez,RS}. Consequently, in this work we refer to the effects of the kinetically dominated period on the evolution of the primordial perturbations as short-lived inflation effects.

Deep in the pre-inflationary period of kinetic dominance, the contribution of the inflaton potential $V(\phi)$ to the energy density $\rho$ should be negligible compared to the kinetic term. Therefore, we directly approximate the evolution of the cosmological background in the period $(\eta_0,\eta_i)$ by restricting eq. \eqref{backGR} to the case $V(\phi)=0$. The resulting dynamics is analytically solvable, with expanding solutions given by
\begin{equation}\label{backin}
a(\eta)=a_0 \sqrt{1+2a_0 H_0 (\eta -\eta_0)},\qquad \rho(\eta)=\rho_0\left[\frac{a_0}{a(\eta)}\right]^{6},
\end{equation}
where $H=\sqrt{8\pi\rho/3}$ is the Hubble parameter, with $H_0=H(\eta_0)$, $a_0=a(\eta_0)$, and $\rho_0=\rho(\eta_0)$. For vanishing inflaton potential, one has $\mathcal{U}_{\mathrm{GR}}=0$, and then the evolution equations for the Mukhanov-Sasaki and tensor perturbations take both the form 
\begin{equation}\label{skin}
v^{(r)\prime\prime}_{\vec{k}}+\left(k^2+s_{\text{kin}}\right)v^{(r)}_{\vec{k}}=0,\qquad s_{\text{kin}}=\frac{1}{4}\left(\eta-\eta_0+\frac{1}{2H_0 a_0}\right)^{-2}.
\end{equation}

According to our hypotheses, the kinetic epoch ends when $\eta=\eta_i$. From this instant on, the potential contribution to the inflaton energy density becomes comparable to the kinetic one, and eventually dominates over it. It is after this transition when the slow-roll inflationary period can start, period in which the cosmological background experiences a quasi-de Sitter expansion. For the purposes of this work, we will simplify the physical evolution and completely neglect any of the transition effects from the kinetically dominated period to the slow-roll inflationary one, as well as any change in the inflaton potential that would produce departures in the expansion from an exact de Sitter behavior during slow-roll.  These simplifications will allow an analytical investigation of the perturbations dynamics.

More specifically, we approximate the evolution of our cosmological background in the period $[\eta_i,\eta_{\mathrm{end}}]$ by Friedmann equations in vacuo with a cosmological constant. The expanding solutions are
\begin{equation}\label{ads}
a(\eta)=\left[a_i^{-1}-H_{\Lambda}(\eta-\eta_i)\right]^{-1}, 
\end{equation}
where $a_i=a(\eta_i)$ and $H_{\Lambda}$ is the constant Hubble parameter in de Sitter. We can single out a unique solution if, at the transition time $\eta_i$, we impose continuity of the scale factor and its first derivative with a solution of the preceding kinetically dominated phase. This fixes $a_i$ and $H_{\Lambda}$ as
\begin{equation}
a_{i}=a_0 \sqrt{1+2a_0H_0 (\eta_{i}-\eta_0)},\qquad H_{\Lambda}=\sqrt{\frac{8\pi}{3}\rho_0}\left[\frac{a_0}{a_i}\right]^{3}.
\end{equation}

On the other hand, on a de Sitter background, the time-depentent mass is given by $-a''/a$ for both the Mukhanov-Sasaki and tensor perturbations. Thus, their dynamical equations in this period are\footnote{Strictly speaking, with our instantaneous transition from the kinetic epoch, the derivative $a''$ would not exist at $\eta_i$. Nevertheless, we can define it as the corresponding limit from the right of its expression for $\eta>\eta_i$.}
\begin{equation}\label{sds}
v^{(r)\prime\prime}_{\vec{k}}+\left(k^2+s_{\text{dS}}\right)v^{(r)}_{\vec{k}}=0,\qquad s_{\text{dS}}=-2H_{\Lambda}^2\left[a_i^{-1}-H_{\Lambda}(\eta-\eta_i)\right]^{-2}.
\end{equation}

So, in total, we approximate the evolution of cosmological perturbations on a single-field inflationary universe with a kinetically dominated pre-inflationary interval by means of the mode equations
\begin{equation}\label{modeqgr}
v^{(r)\prime\prime}_{\vec{k}}+\left(k^2+\tilde{s}_{\text{GR}}\right)v^{(r)}_{\vec{k}}=0,\qquad r=s,t,
\end{equation}
where the approximate time-dependent mass $\tilde{s}_{\text{GR}}$ is piecewise-defined as
\begin{equation}\label{sgrapp}
\tilde{s}_{\text{GR}}=\begin{cases}s_{\text{kin}},\qquad \eta\in(\eta_0,\eta_i),\\ s_{\text{dS}}\qquad \eta\in[\eta_i,\eta_{\mathrm{end}}],\end{cases}
\end{equation}
with $s_{\text{kin}}$ and $s_{\text{dS}}$ respectively given in eqs. \eqref{skin} and \eqref{sds}.

\subsection{Dynamics of gauge invariants: Continuous solutions}

The evolution eqs. \eqref{modeqgr} that we have proposed for the Mukhanov-Sasaki and tensor perturbations present a discontinuity at $\eta=\eta_i$, since we clearly have that $s_{\text{kin}}(\eta_i)\neq s_{\text{dS}}(\eta_i)$. This loss of continuity poses an obstruction to the existence and uniqueness of the solutions in all of $(\eta_0,\eta_{\mathrm{end}}]$. Nonetheless, we can isolatedly study the considered equation restricted to the separate intervals $(\eta_0,\eta_i)$ and $[\eta_i,\eta_{\mathrm{end}}]$ where there is continuity. Then, we can characterize all of the functions that, in addition to being solutions in each of these two intervals, are also continuous with a continuous derivative at $\eta=\eta_i$. These solutions are uniquely determined by their initial conditions. They make promising candidates to approximate the exact dynamics of the perturbations in our considered classical background with inflaton, given the smoothness of the time-dependent mass in the exact model. This is motivated by our expectation that a very localized change on the mass $\tilde{s}_{\text{GR}}$ around $\eta_i$ to turn it into a continous function would not modify too much the solution of the dynamical equation for the perturbative modes. At most, we would expect that this discontinuity (and others of similar nature that we will encounter) might introduce an artificially fast rate of variation of the considered solution around it. This effect might propagate forward and translate into some spuriously generated oscillations, possibly on top of other oscillations associated instead with the choice of vacuum state. If this were the case, we would have to identify those spurious oscillations and consistently remove them, as we will explain later on our discussion.\footnote{For more information on the possible origins of the unwanted behavior that we are referring to, and how to correct it, see the end of section \ref{sec:ii}, the final paragraph of subsection \ref{sec:iv1}, and the remarks in the third paragraph of the concluding section.}

In the interval $(\eta_0,\eta_i)$ the dynamical equation for the Mukhanov-Sasaki and tensor perturbations is given in eq. \eqref{skin}. One can readily check that the general solution $\mu_k$ of this equation is
\begin{equation}\label{mukin}
\mu_k=C_k \sqrt{\frac{\pi y}{4}} H^{(1)}_0 (ky)+D_k \sqrt{\frac{\pi y}{4}} H^{(2)}_0 (ky), \qquad y=\eta-\eta_0+\frac{1}{2H_0 a_0},
\end{equation}
where $C_k$ and $D_k$ are complex integration constants, while $H^{(1)}_{\nu}$ and $H^{(2)}_{\nu}$, with $\nu\in \mathbb{N}$, respectively denote the Hankel functions of the first and second kind \cite{abram}. Notice that the solutions depend only on the norm of the wavevector, $k=|\vec{k}|$. This explains our notation for the subindex of $\mu_k$. Each set of solutions $\mu_k$ for all $\vec{k}$ can be understood as the set of positive-frequency solutions associated with a vacuum state for the scalar and tensor perturbations in the kinetically dominated period if they are normalized as
\begin{equation}\label{normalization}
\mu_k \mu_{k}^{*\prime}-\mu_{k}^{\prime}\mu_k^{*}=i,
\end{equation}
where the symbol $*$ denotes complex conjugation. This normalization condition is equivalent to $|D_k|^2-|C_k|^2=1$. We emphasize that, in the considered time interval, each choice of constants $C_k$ and $D_k$ satisfying this requirement univocally determines a normalized solution $\mu_k$.

On the other hand, for our approximate evolution equations \eqref{sds} in the inflationary period $[\eta_i,\eta_{\mathrm{end}}]$, the general solution $\mu_k$ is well known \cite{mukhanov1}:
\begin{align}\label{muds}
\mu_k=&A_k \frac{e^{ik(\eta-\eta_i-a_{i}^{-1}H_{\Lambda}^{-1})}}{\sqrt{2k}}\left[1+\frac{i}{k(\eta-\eta_i-a_{i}^{-1}H_{\Lambda}^{-1})}\right]\nonumber \\ +&B_k \frac{e^{-ik(\eta-\eta_i-a_{i}^{-1}H_{\Lambda}^{-1})}}{\sqrt{2k}}\left[1-\frac{i}{k(\eta-\eta_i-a_{i}^{-1}H_{\Lambda}^{-1})}\right],
\end{align}
where $A_k$ and $B_k$ are complex integration constants. This solution is normalized according to eq. \eqref{normalization} if $|B_k|^2-|A_k|^2=1$. Once again, we notice that any normalized choice of integration constants, for all $\vec{k}$, specifies a set of positive-frequency solutions for the cosmological perturbations in the de Sitter regime. 

Using the preceeding information, we can uniquely characterize any physically sensible family of mode solutions for the Mukhanov-Sasaki and tensor perturbations that is defined in the whole time interval $(\eta_0,\eta_{\text{end}}]$ as follows. Starting with the general expressions \eqref{mukin} and \eqref{muds} for the (normalized) solutions in, respectively, $(\eta_0,\eta_i)$ and $[\eta_i,\eta_{\mathrm{end}}]$, if we impose continuity up to the first derivative at $\eta=\eta_i$, we obtain after a direct computation that
\begin{align}\label{Ak}
A_k=&\frac{e^{ik/(a_i H_{\Lambda})}}{4}\sqrt{\frac{k\pi}{a_i H_{\Lambda}}}\bigg\{C_k\bigg[H_0^{(1)}\left(\frac{k}{2a_i H_{\Lambda}}\right)-\left(\frac{a_i H_{\Lambda}}{k}-i\right)H_1^{(1)}\left(\frac{k}{2a_i H_{\Lambda}}\right)\bigg]\nonumber \\+& D_k\bigg[H_0^{(2)}\left(\frac{k}{2a_i H_{\Lambda}}\right)-\left(\frac{a_i H_{\Lambda}}{k}-i\right)H_1^{(2)}\left(\frac{k}{2a_i H_{\Lambda}}\right)\bigg]\bigg\},
\end{align}
\begin{align}\label{Bk}
B_k=&\frac{e^{-ik/(a_i H_{\Lambda})}}{4}\sqrt{\frac{k\pi}{a_i H_{\Lambda}}}\bigg\{C_k\bigg[H_0^{(1)}\left(\frac{k}{2a_i H_{\Lambda}}\right)-\left(\frac{a_i H_{\Lambda}}{k}+i\right)H_1^{(1)}\left(\frac{k}{2a_i H_{\Lambda}}\right)\bigg]\nonumber \\+& D_k\bigg[H_0^{(2)}\left(\frac{k}{2a_i H_{\Lambda}}\right)-\left(\frac{a_i H_{\Lambda}}{k}+i\right)H_1^{(2)}\left(\frac{k}{2a_i H_{\Lambda}}\right)\bigg]\bigg\}.
\end{align}
This relation between the integration constants in each of the two subintervals is invertible. Furthermore, if $C_k$ and $D_k$ are normalized as indicated above, then the corresponding coefficients $A_k$ and $B_k$ are also normalized, and vice-versa. It follows that any mode solution of our dynamical eqs. \eqref{modeqgr} defined in $(\eta_0,\eta_{\text{end}}]$ that is continuous and has a continuous derivative is fixed once one provides a choice of (normalized) integration constants $C_k$ and $D_k$ (or alternatively their counterpart $A_k$ and $B_k$). We henceforth restrict our attention to solutions of this type, and regard any choice of integration constants of this kind as a choice of vacuum state or, equivalently, of initial conditions for the primordial perturbations.

\subsection{Primordial power spectrum: Non-oscillating behavior}

In single-field inflationary cosmology, the standard definition of the primordial power spectrum for scalar and tensor perturbations is respectively\footnote{The time of evaluation $\eta_{\text{end}}$ can also be set a few e-folds after all of the Fourier scales $k$ that are observable nowadays have exited the Hubble horizon, since their (rescaled) evolution freezes then.} \cite{langlois}
\begin{equation}
\label{primordialpower}
\mathcal{P_{\mathcal{R}}}(k)=\frac{k^3}{2\pi^2}  \frac{|\mu_k (\eta_{\rm end})|^2} {z^2(\eta_{\rm end})}, \quad \quad   \mathcal{P_{\mathcal{T}}}(k)=\frac{32 k^3}{\pi}  \frac{|\mu_k (\eta_{\rm end})|^2} {a^2(\eta_{\rm end})},
\end{equation}
where $\mu_k$ are solutions of eq. \eqref{modexactgr} with the normalization \eqref{normalization}. On the other hand, we recall that observationally admissible inflationary epochs must last at least around $60$ e-folds \cite{liddlef}. Taking this fact into account, in a situation where one considers an exact de Sitter inflation, it is reasonable to approximate the primordial power spectra (at dominant order) as
\begin{equation}
\mathcal{P}(k)=\frac{k^3}{2\pi^2}\lim_{a\rightarrow\infty}\left(\frac{|\mu_k|^2}{a^2}\right),
\end{equation}
up to a constant factor in the case of tensor perturbations that is irrelevant for the purposes of this work. Employing eq. \eqref{ads} for the scale factor in de Sitter, and eq. \eqref{muds} for the general solution of the Mukhanov-Sasaki and tensor perturbations in this regime, one concludes that
\begin{equation}\label{powerdesitter}
\mathcal{P}(k)=\frac{H_{\Lambda}^2}{4\pi^2}|B_k-A_k|^2.
\end{equation}
This is the formula that we will use to estimate the power spectra in our approximate description of primordial fluctuations over a single-field inflationary cosmology with a non-trivial kinetic pre-inflationary epoch.

It is clear that the specific form of the primordial power spectrum \eqref{powerdesitter} depends critically on the choice of (normalized) integration constants $A_k$ and $B_k$ (or, equivalently, of $C_k$ and $D_k$) that fix the mode solutions for the perturbations. In other words, the behavior of the spectrum is strongly affected by the choice of vacuum state for the perturbations. For example, one can proceed as in standard inflationary models and choose the Bunch-Davies state \cite{BD}, which is the preferred one in a genuine cosmological de Sitter spacetime. Such vacuum state is characterized by $|B_k|=1$ and $A_k=0$, and yields the well-known scale-invariant power spectrum \cite{mukhanov1}. This choice can be thought of as natural if the perturbations with observationally relevant Fourier scales $k$ were not significantly affected by the pre-inflationary dynamics, as it would happen if slow-roll inflation lasted long enough so as to wash away any pre-inflationary effect. However, if this is not the case, the Bunch-Davies vacuum loses its privileged status and the question of which is the natural state for the cosmological fluctuations becomes especially important in order to extract predictions.

In order to get a preliminary hint of the qualitative behavior that can result for the power spectra from eq. \eqref{powerdesitter}, it is convenient to re-express this formula as
\begin{equation}\label{cosineq}
\mathcal{P}(k)=\frac{H_{\Lambda}^2}{4\pi^2}\bigg[|B_k|^2+|A_k|^2-2|A_k||B_k|\cos\left(\theta_k^A-\theta_k^B\right)\bigg]
\end{equation}
where $\theta_k^A$ and $\theta_k^B$ are the respective phases of $A_k$ and $B_k$. This expression is well suited to understand why, in certain cases, the primordial power spectrum is going to display an oscillatory behavior in the scale $k$, and how to avoid these oscillations. Let us consider all choices of integration constants $A_k$ and $B_k$ such that their complex norms are non-oscillating functions of $k$. Then, it is clear that any $k$-dependent relative phase between these constants results into a primordial power spectrum that oscillates in $k$, unless $|A_k|$ is completely negligible compared to $|B_k|$ (we recall that the normalization condition implies that $|B_k|>|A_k|$). It is also worth noticing that, for many choices of this type, the resulting spectrum can display a mixed behavior, in which there are oscillations in those intervals of $k$ where $|A_k|$ and $|B_k|$ are of comparable order, and no oscillations for all $k$ such that $|A_k|\ll |B_k|$ or the difference of phase between $A_k$ and $B_k$ hardly varies.

In section \ref{sec:iv} we will argue in favor of a state for the primordial perturbations leading to a power spectrum without those oscillations in $k$. One of the reasons for this preference is that those oscillations typically involve an enhancement of power, which can be regarded as artificial. For future use, it is worth explaining a straightforward method for eliminating the oscillations in the spectrum for integration constants $A_k$ and $B_k$ with non-oscillatory norms and a $k$-dependent dephasing. Actually, the resulting power spectrum captures the main information that gets amplified in the presence of the oscillations, as we will see. One simply has to consider a redefinition of the integration constants of the form
\begin{align}
A_k\rightarrow \tilde{A}_k=|A_k|,\qquad B_k\rightarrow \tilde{B}_k=|B_k|.
\end{align}
In fact, this is a Bogoliubov transformation, in the sense that it leads to new normalized integration constants. The primordial power spectrum for these new tilded constants is
\begin{align}
\tilde{\mathcal{P}}(k)=\frac{H_{\Lambda}^2}{4\pi^2}\left(|B_k|-|A_k|\right)^2.
\end{align}
Clearly, this spectrum does not display fast oscillations. Instead, as a function of $k$, it follows the trajectory of the minima of the oscillations of the original spectrum. In this sense, it indeed maintains the physically relevant information of the original spectrum once the amplifying oscillations have been removed.

The procedure that we have just outlined is crucial in the derivation of the spectrum, because it removes the oscillations from it in a very specific way. Its physical implications are very important and in consonance with the proposal of adopting a non-oscillating vacuum state in the calculation of the spectrum  that favors the suppression of power. In this sense, it is worth noticing that the suggested procedure is not the unique way in which one can get rid of an oscillatory behavior. Indeed, taking a look at eq. \eqref{cosineq} one immediately realizes that, for non-oscillatory $|A_k|$ and $|B_k|$, it suffices to redefine the dephasing between $A_k$ and $B_k$ so that the cosine in this formula becomes a $k$-independent number. The strategy that we have just introduced simply corresponds to setting this cosine equal to one, namely to completely remove the difference of phase. This procedure is the only one that captures the minimum values of the power spectrum that get amplified in the presence of oscillations. In Appendix \ref{app0} we argue from an analytical perspective how, starting from a suitable initial state for the perturbations, one may regard the possible dephasing occuring between $A_k$ and $B_k$ as a consequence of the matching between positive-frequency solutions corresponding to different epochs, which we perform owing to the discontinuities that our approximations introduce in the evolution of the perturbations. Therefore, we view our strategy of removing this dephasing as a well-motivated and preferred course of action to eliminate the spurious oscillations appearing in the power spectra that we compute in this work.

\section{Approximate mass for the perturbations in hybrid LQC}
\label{sec:iii}

Going one step beyond the classical inflationary cosmologies previously discussed, in this section we consider their counterparts in the framework of effective LQC. Let us first introduce the model in hybrid LQC that we will later approximate in order to investigate the evolution of the cosmological perturbations on a background that corresponds to a solution of effective (homogeneous and isotropic) LQC. Without backreaction and in the presence of an inflaton $\phi$, these background equations are given in conformal time by \cite{APS,jorma,hybr-pred}
\begin{equation}\label{LQC}
\left(\frac{a'}{a}\right)^{2}=\frac{8\pi }{3}a^2\rho \left(1-\frac{\rho}{\rho_{c}}\right), \qquad \frac{a''}{a}=\frac{4\pi }{3}a^2\rho\left(1+2\frac{\rho}{\rho_{c}}\right)-4\pi  a^2 P\left(1-2\frac{\rho}{\rho_{c}}\right).
\end{equation}
The energy density $\rho$ and pressure $P$ of the inflaton are defined in eq. \eqref{rhop}, and $\rho_c\simeq 0.4092$ is the maximum value that the energy density can take in effective LQC. The bounce occurs at the moment when this value is reached. These equations are a way of modelling the trajectories of the peaks of certain Gaussian states in homogeneous and isotropic LQC. When the energy density $\rho$ falls to be much smaller than $\rho_c$, they are simply the classical Friedmann equations considered in section \ref{sec:ii}. 

In the hybrid approach to the quantization of cosmological perturbations in LQC, the dynamical equations for the Mukhanov-Sasaki and tensor modes in the considered effective regime adopt the form \cite{hyb-pert4,hybr-ten}
\begin{equation}\label{modexactlqc}
v^{(r)\prime\prime}_{\vec{k}}+\left[k^2+s^{(r)}_h\right]v^{(r)}_{\vec{k}}=0,\qquad r=t,s,
\end{equation}
where \cite{mass}
\begin{equation}
s^{(t)}_h=-\frac{4\pi }{3}a^2(\rho-3P),\qquad s^{(s)}_h= s^{(t)}_h+\mathcal{U}_{h},
\end{equation}
and $\mathcal{U}_{h}$ is a linear combination (with time-dependent coefficients) of the potential $V(\phi)$, its square, and its first and second derivatives with respect to $\phi$. In practice, the time-dependent masses $s^{(r)}_h$ ($r=t,s$) become indistinguishable from the classical masses \eqref{massgr} as soon as the background solution enters the low-energy regimes with $\rho\ll \rho_c$. However, the difference between these masses is considerable in the regions surrounding the loop quantum bounce.

\subsection{Approximate background solutions}

Equations \eqref{LQC} in effective LQC lead to solutions with a bounce and that display a classical behavior at low matter energy densities. The specific evolution of the solution after the bounce depends on the initial conditions chosen there. Here, we are particularly interested in solutions with phenomenological relevance. Namely, we will focus our attention on effective LQC solutions such that their perturbations with the largest wavelengths that would be observable today in the CMB may have been affected by loop quantum effects in the neighbourhood of the bounce. Roughly speaking, this means that the lowest values of $k$ corresponding to the observable power spectrum are such that $k^2$ is close to or slightly smaller than $s_h^{(r)}$ in the surroundings of the bounce of the considered background. 

This type of effective LQC solutions have been widely studied over the last decade. Remarkably, they all display the following qualitative behavior. The inflaton energy density is dominated by its kinetic contribution at the bounce and, shortly after, a classical period of deccelerated expansion follows, leading finally to a slow-roll inflationary period that is short-lived \cite{Ivan,hybr-pred,Universe,AGvacio2}. In the case of a quadratic potential for the inflaton, these solutions correspond to choices of inflaton mass of the order $m\sim 10^{-6}$ and an initial value of $\phi$ at the bounce of order one (in Planck units). On the one hand, these values for $m$ are derived by requiring a number of e-folds of slow-roll inflation that can be compatible with the minimum allowed by observations. Remarkably, for all initial values of $\phi$ of order one or greater, this restriction on $m$ guarantees that the scales $k$ that are significantly larger than the spacetime curvature at the bounce lie in the observationally (quasi-)scale-invariant part of the power spectrum. On the other hand, the restriction on the initial value of $\phi$ at the bounce follows from demanding that inflation does not last long enough so as to shift the smallest observable scales $k$ to be much greater than the spacetime curvature at the bounce.\footnote{The restriction on the initial datum for $\phi$ was first obtained numerically after fixing a specific vacuum state for the perturbations \cite{Ivan}. Nonetheless, we consider reasonable to expect that this numerical restriction does not change much if it is reformulated independently of the choice of vacuum, by asking that the smallest observational scales $k$ are close to the curvature at the bounce \cite{Universe,AGvacio2,AshNe}.} For this type of effective LQC solutions, the inflaton energy density at the onset of inflation ranges in the interval $[10^{-9},10^{-12}]$ \cite{Universe}.

Restricting our discussion to these solutions of phenomenological interest, it seems reasonable that, during a large part of the pre-inflationary epoch, one can approximate the dynamics by setting $V(\phi)=0$ in  eq. \eqref{LQC}. The evolution  can then be solved analytically in the proper time $t$, obtaining \cite{wang2}
\begin{equation}\label{kineticlqc}
\dot{\phi}(t)=\pm\sqrt{2\rho_c}a^{-3}(t),\qquad a(t)=\left(1+24\pi\rho_c t^2\right)^{1/6}.
\end{equation}
Here, we have arbitrarily set $t=0$ at the bounce, and we have set the global scale of distances such that $a(t=0)=1$, without loss of generality.\footnote{In standard cosmology, one usually sets the present value of the scale factor equal to one. Choosing instead the value at the bounce shifts proportionally all the Fourier scales $k$ in the observable window with respect to those reported by e.g. the Planck mission.} We notice that the proper time $t$ is related to the conformal one by 
\begin{equation}\label{etat}
\eta -\eta_B={}_2 F_1\left(\frac{1}{6},\frac{1}{2};\frac{3}{2};-24\pi\rho_c t^2\right)t,
\end{equation}
where $\eta_B$ is the value of $\eta$ at the bounce and ${}_2 F_1$ is the Gauss hypergeometric function \cite{abram}. In this situation with negligible inflaton potential, one can check that the dynamical equations for both the scalar and tensor perturbations have the form
\begin{equation}\label{modekinlqc}
v^{(r)\prime\prime}_{\vec{k}}+\left(k^2+s^{\text{kin}}_h\right)v^{(r)}_{\vec{k}}=0,\qquad s^{\text{kin}}_h=\frac{8\pi\rho_c}{3}\left(1+24\pi\rho_c t^2\right)^{-2/3}.
\end{equation}
As we have already commented, the effective LQC trajectories rapidly adopt a classical behavior after the bounce occurs. Focusing on solutions of the considered type, it is then reasonable to expect that the analysis and approximations developed in section \ref{sec:ii} can be applied now as well, starting at a certain instant $\eta_0$ in the kinetically dominated regime. In this way, we will have at our disposal a completely analytical description, at least approximate, of our effective LQC background. 

More specifically, the initial conditions for eq. \eqref{backin} at that time $\eta_0$ (in the kinetic epoch) that can approximate well the effective LQC evolution of our background towards the future are
\begin{equation}\label{inickin}
a_0=\left(1+24\pi\rho_c t_0^2\right)^{1/6},\qquad \rho_0=\rho_c a_0^{-6},
\end{equation}
where $t_0=t(\eta_0)$ is calculated using eq. \eqref{etat}. The accuracy
of this approximation depends on the choice of time $\eta_0$ for the matching of initial conditions using the LQC data. In particular, since our interest here is the study of the propagation of cosmological perturbations, an acceptable choice of $\eta_0$ in the kinetically dominated region is one for which the difference between the hybrid LQC mass $s^{\text{kin}}_h$ and the classical mass $s_{\text{kin}}$ becomes negligible. This latter mass is obtained from eq. \eqref{skin}, evaluated at the initial conditions \eqref{inickin} inherited from effective LQC. According to our comments, it is then convenient to introduce the relative difference
\begin{align}
2\frac{|s_{\text{kin}} -s^{\text{kin}}_h|}{(s_{\text{kin}}+s^{\text{kin}}_h)}	
\end{align}
which measures the relative error made when one approximates the hybrid LQC mass by its classical counterpart in the kinetic epoch. This difference can be computed analytically as a function of time, for different choices of $\eta_0$. The results show that, for $t_0\gtrsim 0.4$, the error committed grows at most up to a $3\%$ immediately after the initial time $t_0$ (instant at which both masses coincide), and becomes negligible at large $t>t_0$. However, the maximum of this error rapidly increases if one chooses smaller values of $t_0$.

\subsection{P\"oschl-Teller approximation}

Even with the approximated analytical description of the effective LQC background at our disposal, we are not able to solve exactly the mode equation \eqref{modekinlqc} for the Mukhanov-Sasaki and tensor perturbations. This situation calls for an additional approximation to the time-dependent mass $s^{\text{kin}}_h$ near the bounce, that allows us to find explicit solutions for the cosmological perturbations. 

\begin{figure}
\centering
\includegraphics[width=7.5 cm]{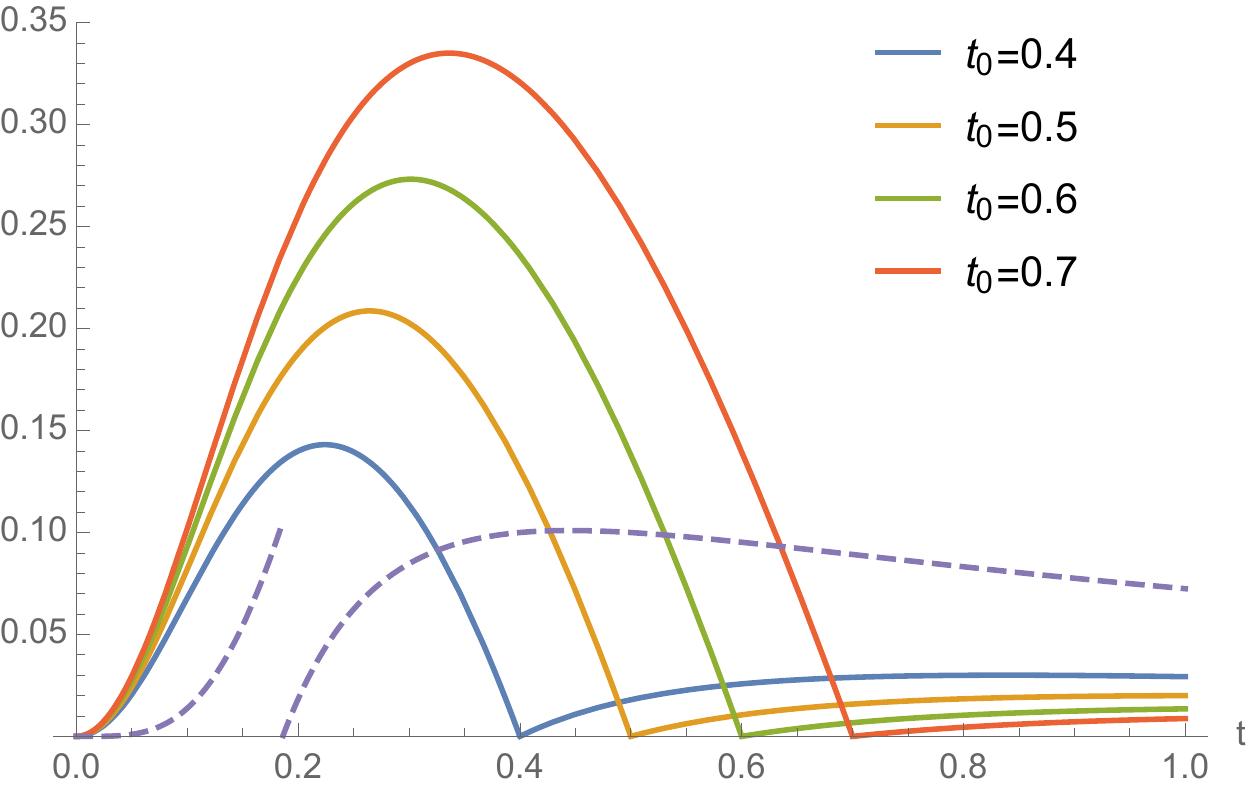}\includegraphics[width=7.5 cm]{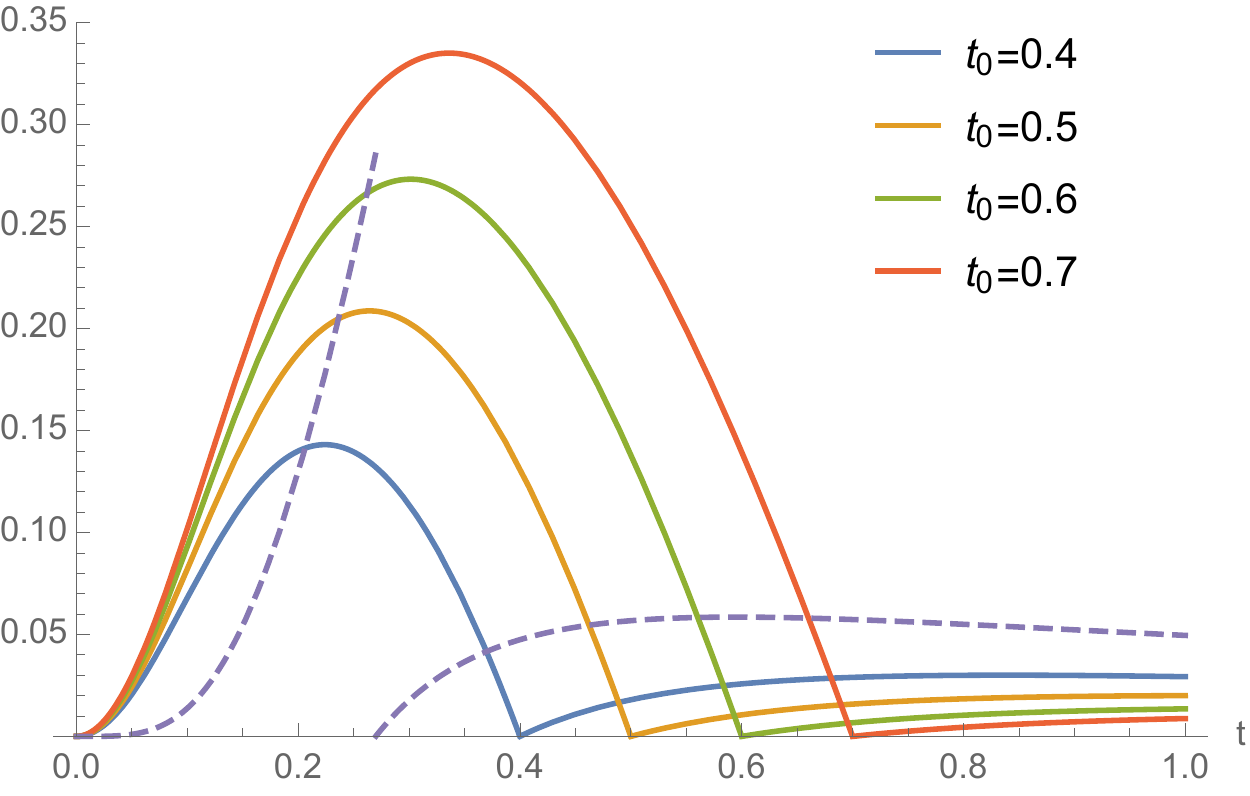}
\caption{The relative difference $\mathrm{Err}$ as a function of the proper time $t$, for several choices of $t_0\geq 0.4$. Dashed lines: The corresponding function $\mathrm{Err}$ defined using the P\"oschl-Teller potential with $\alpha=\sqrt{16\pi\rho_c}$, for $t_0\simeq 0.19$ (left panel) and $t_0\simeq 0.27$ (right panel). }
\label{fig1} 
\end{figure}

In this respect, refs. \cite{wang,wang2} suggested that the behavior of the mass $s^{\text{kin}}_h$ in the vicinity of the bounce closely resembles that of a P\"oschl-Teller potential of the form
\begin{equation}\label{PT}
s_{\mathrm{PT}}=\frac{U_0}{\cosh^{2}{[\alpha(\eta-\eta_B)]}}.
\end{equation}
The works that have used this expression to approximate the mass in hybrid LQC have so far chosen the values $U_0=8\pi\rho_c/3$ and $\alpha=\sqrt{16\pi\rho_c}$ for the respective peak and width of the potential \cite{wang}. These values are obtained by demanding that the functions $s_{\mathrm{PT}}$ and $s^{\text{kin}}_h$ coincide at the bounce, up to the second time derivative (included). However, if one compares this choice of P\"oschl-Teller mass with $s^{\text{kin}}_h$, one realizes that the relative error rapidly grows beyond a $20\%$ for $t\gtrsim0.25$ (see the dashed lines in figure \ref{fig1}). Then, according to our discussion above, the interval of time for which this P\"oschl-Teller mass can be considered a fairly good approximation is too narrow to include admissible matching points with the classical kinetically dominated background.

With the aim of modelling a more realistic transition to the classical regime, we can improve the P\"oschl-Teller approximation of the hybrid LQC mass near the bounce by changing its width to values other than $\alpha=\sqrt{16\pi\rho_c}$. More concretely, we demand that $s_{\mathrm{PT}}$ coincide with $s^{\text{kin}}_h$ at the bounce and with $s_{\text{kin}}$ at the time $\eta_0$ where we connect the effective LQC evolution with the classical kinetic epoch. This requirement fixes the peak and width of the P\"oschl-Teller potential in the following way:
\begin{equation}\label{alphau}
 U_0=\frac{8\pi\rho_c}{3}, \qquad \alpha=\frac{\text{arcosh}(a_0^2)}{(\eta_0-\eta_B)},
\end{equation}
as one can check using eqs. \eqref{skin}, \eqref{modekinlqc}, and \eqref{inickin}. Therefore, we can try and approximate the evolution of the primordial cosmological perturbations on the effective LQC background with short-lived inflation as
\begin{equation}\label{modeqlqc}
v^{(r)\prime\prime}_{\vec{k}}+\left(k^2+\tilde{s}_{h}\right)v^{(r)}_{\vec{k}}=0,\qquad r=s,t,
\end{equation}
where the time-dependent mass $\tilde{s}_{h}$ is
\begin{equation}
\tilde{s}_{h}=\begin{cases}U_0\cosh^{-2}{[\alpha(\eta-\eta_B)]},\qquad \eta\in[\eta_B,\eta_0],\\ \tilde{s}_{\text{GR}}\qquad \eta\in(\eta_0,\eta_{\mathrm{end}}],\end{cases}
\end{equation}
and $\alpha$ and $U_0$ are given in eq. \eqref{alphau}. Recall that $\tilde{s}_{\text{GR}}$ is our approximation to the classical mass [see eq. \eqref{sgrapp}].

Let us notice that the choice of parameters for the P\"oschl-Teller potential made in eq. \eqref{alphau} depends on the instant $\eta_0$ where we place the transition to a classical universe. In order to elucidate which values of $\eta_0$ are acceptable, it is convenient to define the relative difference
\begin{equation}
\mathrm{Err}=2\frac{|\tilde{s}_{h} -s^{\text{kin}}_h|}{(\tilde{s}_{h}+s^{\text{kin}}_h)},
\end{equation}
which measures the relative error made when one approximates the hybrid LQC mass in the kinetic epoch as we have suggested. Figure \ref{fig1} shows this error as a function of time, for several choices of $t_0\geq 0.4$. For completeness, this figure also displays the error that we would have made if we had defined $\tilde{s}_{h}$ using the P\"oschl-Teller potential with $\alpha=\sqrt{16\pi\rho_c}$. This proves that our proposed parameters \eqref{alphau} provide an overall improvement of the existing approximations of the hybrid LQC mass employing a P\"oschl-Teller term \cite{wang}, without compromising its classical behavior. Furthermore, inspecting figure \ref{fig1}, one can argue that the choices of transition time $t_0$ that optimize our approximation lie around $t_0\simeq 0.4$.

\subsection{Approximate dynamics of gauge invariants: Continuous solutions}

The dynamical eq. \eqref{modeqlqc}, introduced to approximate the evolution of the perturbations in hybrid LQC for effective backgrounds with short-lived inflation, can be solved analytically in the two time intervals $[\eta_B,\eta_0]$ and $(\eta_0,\eta_{\mathrm{end}}]$. For this last interval, which corresponds to the classical period, the general solution with continuity up to the first derivative has been characterized in section \ref{sec:ii}. On the other hand, in the interval $[\eta_B,\eta_0]$ that describes the vicinity of the bounce, the perturbations evolve according to
\begin{equation}
v^{(r)\prime\prime}_{\vec{k}}+\left(k^2+s_{\mathrm{PT}}\right)v^{(r)}_{\vec{k}}=0,\qquad r=s,t,	
\end{equation}
where $s_{\mathrm{PT}}$ is the P\"oschl-Teller mass \eqref{PT} with parameters \eqref{alphau}. One can check that its general solution $\mu_k$ is
\begin{align}\label{mupt}
\mu_k=&M_k[x(1-x)]^{-ik/2\alpha}{}_2 F_1\left(b_1^k,b_2^k;b_3^k;x\right)\nonumber \\+&N_k x^{ik/2\alpha}(1-x)^{-ik/2\alpha}{}_2 F_1\left(b_1^k-b_3^k+1,b_2^k-b_3^k+1;2-b_3^k;x\right),
\end{align}
where $M_k$ and $N_k$ are integration constants, and we have defined the time variable
\begin{equation}\label{xtime}
x=\left[1+e^{-2\alpha(\eta-\eta_B)}\right]^{-1}.
\end{equation}
In addition, the parameters of the hypergeometric function are 
\begin{align}\label{a3}
b_3^k=1-\frac{ik}{\alpha},
\end{align}
\begin{align}\label{a12}
b_1^k=\frac{1}{2}\left(1+\sqrt{1+\frac{32\pi\rho_c}{3\alpha^2}}\right)-\frac{ik}{\alpha},\qquad b_2^k=\frac{1}{2}\left(1-\sqrt{1+\frac{32\pi\rho_c}{3\alpha^2}}\right)-\frac{ik}{\alpha}.
\end{align}
This solution can be normalized according to eq. \eqref{normalization} with a suitable restriction on the constants $M_k$ and $N_k$, which we need not show here.

At the instant $\eta_0$, the approximate mass function $\tilde{s}_h$ is continuous by construction. Then, each choice of integration constants $M_k$ and $N_k$, namely of a solution in the surrounding of the bounce, selects a unique solution \eqref{mukin} in the kinetic epoch. The constants $C_k$ and $D_k$ associated with this solution are obtained by demanding its continuity and that of its derivative at $\eta_0$. They are given by
\begin{equation}\label{Ck}
C_k=\left[H_0^{(1)}\left(\frac{k}{2H_0 a_0}\right)\right]^{-1}\bigg[\sqrt{\frac{8H_0 a_0}{\pi}}\mu_k(\eta_0)-D_k H_0^{(2)}\left(\frac{k}{2H_0 a_0}\right)\bigg],
\end{equation}
\begin{align}\label{Dk}
D_k=&i\sqrt{\frac{\pi}{8H_0 a_0}}\bigg[kH_{1}^{(1)}\left(\frac{k}{2H_0 a_0}\right)\mu_k(\eta_0)-H_{0}^{(1)}\left(\frac{k}{2H_0 a_0}\right)H_0 a_0\mu_k(\eta_0)\nonumber \\&+H_{0}^{(1)}\left(\frac{k}{2H_0 a_0}\right)\mu'_k(\eta_0)\bigg].
\end{align}
As we explained in section \ref{sec:ii}, any choice of these constants fixes in turn a solution in the de Sitter period provided that it is continuous and has a continuous derivative at $\eta_i$, by means of eqs. \eqref{Ak} and \eqref{Bk}. 

In summary, if we restrict our attention to solutions of our approximate equation \eqref{modeqlqc} that are $C^{1}$ at the matching points, each choice of integration constants $M_k$ and $N_k$ in the vicinity of the bounce leads to a unique solution throughout the whole pre-inflationary and inflationary periods, solution that can be calculated explicitly. Alternatively, the same happens if one chooses the pair of constants $C_k$ and $D_k$ in the kinetic epoch, or $A_k$ and $B_k$ in the de Sitter period. The vacuum state associated with any of such choices for all $\vec{k}$ leads to a primordial power spectrum \eqref{powerdesitter}, with properties that we can investigate analytically.

\section{Short-lived inflation vs. loop quantum bounce effects}
\label{sec:iv}

In what follows, we will use the approximations introduced in sections \ref{sec:ii} and \ref{sec:iii} for an analytical study of the differences arising in the primordial power spectra from the evolution of the perturbations on a purely classical universe, on the one hand, and on a hybrid LQC background, on the other hand, both with a period of short-lived inflation. For a robust comparison, the first and most important step that we need to take is to establish a common criterion for the selection of a vacuum state of the perturbations, on any of the considered backgrounds. 

\subsection{Choice of vacuum state}
\label{sec:iv1}

As we discussed in section \ref{sec:ii}, in any situation where a pre-inflationary epoch may have affected the perturbations with scales $k$ that are observable nowadays, the arguments leading to the Bunch-Davies state lose their strength. Instead, one should develop new criteria that distinguish a preferred quantum state for the perturbations, taking into account the pre-inflationary dynamics. In the lack of enough symmetries in the background spacetime that can serve to naturally select such a state, several proposals have emerged over the last decade in the context of LQC \cite{Ivan,no,AGvacio2,wang,wang2,AANv,AGvacio1,RMJ}. A notorious example is the consideration of states constructed with the imposition of (low-order) adiabatic initial conditions \cite{adiabatic1,adiabatic2}. These conditions are usually set at the bounce, or at some instant in the distant past from it. Other more recent proposals involve the minimization of the fields uncertainty relations in the vicinity of the bounce \cite{AGvacio1,AGvacio2}, or of the smeared time variation of the mode solutions throughout a good portion of the evolution from the bounce to the onset of inflation \cite{no}.

All of these proposals to select a vacuum state heavily depend on the choice of time to set the initial conditions, and/or require numerical techniques to determine them. Furthermore, the numerical evolution of the associated mode solutions typically leads to primordial power spectra that are highly oscillating with respect to the scale $k$ \cite{Ivan,hybr-pred,Universe,AGvacio2,RMJ}. In order to extract predictions from the theory, these oscillations are often averaged out. In recent years, concerns have been raised about whether these oscillations are physically acceptable, or they rather should be interpreted as an indication that the corresponding state of the perturbations is not a natural vacuum in the theory \cite{no,Universe,NO-analy}. For example, the Bunch-Davies vacuum does not present such oscillations, not even when quasi-de Sitter effects of the inflationary background destroy its scale-invariance. This state is undoubtedly a preferred one when the attention is restricted to slow-roll inflation, and it provides accurate tested predictions for most observable scales $k$. Moreover, oscillations in the power spectrum can be thought of as the remnants of a past evolution of the mode solutions with fast oscillations in time. Such oscillatory dynamics can easily erase relevant information about the effects of the earliest epochs of the pre-inflationary Universe on the perturbations. Finally, oscillations in the primordial spectrum have been seen to often pump a somewhat artificial amplification of power, which in some cases can even be in slight tension with the CMB observations \cite{Ivan,Universe}.

Let us recall that the aim of this work is to develop manageable tools to investigate the differences between classical (short-lived inflationary) effects and LQC effects on the primordial power spectrum. From this perspective, it would be especially desirable to characterize the vacuum state {\emph {analytically}} by means of a criterion such that it is motivated by fundamental arguments and can potentially lead to a power spectrum with non-oscillating properties. In this respect, a proposal has been recently put forward in hybrid LQC which seems promising to address both requirements \cite{NO-analy,msdiag,ms-corr}:

\begin{itemize} \item[i)] It can be directly related \cite{ms-corr} to the vacuum state associated with a set of annihilation and creation operators that possess appealing physical properties: their evolution in hybrid quantum cosmology is dictated by a diagonal Hamiltonian, in which the self-interacting parts have been eliminated order by order in the local ultraviolet regime $k\rightarrow\infty$ \cite{msdiag}. It is worth noticing that the procedure leading to this vacuum state is very different from an instantaneous diagonalization of the Hamiltonian.
	
\item[ii)] The norm of the mode solutions for the Mukhanov-Sasaki and tensor perturbations corresponding to this vacuum state in the effective LQC regimes minimize the scale and time-dependent oscillations in the ultraviolet regime $k\rightarrow\infty$ \cite{NO-analy}.
\end{itemize}

The aforementioned state is characterized by the mode solutions\footnote{The lower end of the integration interval in eq. \eqref{muh} is set equal to $\eta_0$ for convenience. The solutions for other choices only differ in global constant phases, irrelevant for the power spectrum.}
\begin{equation}\label{muh}
\mu_k=\sqrt{-\frac{1}{2\text{Im}(h_k)}}e^{i\int_{\eta_0}^\eta d\tilde\eta \text{Im}(h_k)(\tilde{\eta})}	
\end{equation}
for all $\vec{k}$, where $h_k$ is a solution of the differential equation
\begin{equation}
h_{k}'=k^2+s_h^{(r)}+h_k^2
\end{equation}
that displays the following asymptotic expansion
\begin{equation}\label{asymph}
kh_k^{-1}\sim i\left[1-\frac{1}{2k^2}\sum_{n=0}^{\infty}\left(\frac{-i}{2k}\right)^{n}\gamma_n \right].
\end{equation}
Here, $\gamma_n$ are $k$-independent functions of time that are fixed by the relations
\begin{equation}\label{recursion}
\gamma_{0}=s_h^{(r)},\qquad \gamma_{n+1}=-\gamma_{n}'+4s_h^{(r)} \left[\gamma_{n-1}+\sum_{m=0}^{n-3}\gamma_m \gamma_{n-(m+3)}\right]-\sum_{m=0}^{n-1}\gamma_m \gamma_{n-(m+1)}.
\end{equation}

Although this proposal of vacuum state was first motivated and introduced in the context of hybrid LQC, one can straightforwardly generalize the formulas that define its mode solutions to other time-dependent masses different from $s_h^{(r)}$. In particular, the proposal selects the Bunch-Davies vacuum if the mass of the primordial perturbations becomes the mass associated with a de Sitter spacetime in GR. However, in the case of the time-dependent masses $\tilde{s}_{\mathrm{GR}}$ and $\tilde{s}_{h}$, with which we respectively approximate the dynamics of the perturbations on classical and LQC backgrounds with short-lived inflation, the above formulas for the selection of a vacuum cannot be directly applied. Actually, we recall that these masses have a discontinuity at the beginning of inflation, at $\eta_i$. In addition, $\tilde{s}_{h}$ is not smooth at $\eta_0$. In spite of these obstructions, the formulas are  nonetheless applicable in each of the pre-inflationary periods $(\eta_0,\eta_i)$ and $[\eta_B,\eta_0]$.\footnote{To use eq. \eqref{recursion} at the right end of the interval $[\eta_B,\eta_0]$, we define all of the derivatives of the mass $\tilde{s}_{h}$ as the corresponding limits from the left.} These periods can be viewed as the earliest intervals of our GR and LQC cosmologies, respectively (with a different suitable choice of $\eta_0$ in each case). Thus, we can and will use the formulas in these intervals to characterize, with the same criteria, the respective preferred sets of mode solutions on the classical and the effective LQC backgrounds. 

Following this procedure, however, it is reasonable to expect that the vacuum state resulting for each model (relativistic or within effective LQC) is not well adapted to the full dynamics of the background. This is so because of the discontinuities that our approximations introduce in the effective masses $\tilde{s}_{\mathrm{GR}}$ and $\tilde{s}_{h}$. Indeed, the iterative formula \eqref{recursion} collects the information about the global behavior of the mass function, provided that it is smooth, throughout the whole cosmological evolution. If this mass displays discontinuities, as it is the case with our analytical approximations, the only information that it can collect is about the particular smooth interval in which it is applied. Thus, a priori, the vacuum state  is only adapted to the dynamical behavior of the background in the considered interval. In this sense, and as anticipated in section \ref{sec:ii}, we expect that there will appear oscillations in the final power spectrum associated with this state, given that, when encountering any of the transitional discontinuities, the mode solutions that are adapted to the preceeding period of cosmological evolution have to rapidly accomodate to an instantaneous and sudden change. In view of this limitation of our treatment, we will argue in favor of the application of formula \eqref{recursion} in the earliest of the smooth intervals of evolution for each of the considered cosmological models. Then, with the resulting primordial spectra at our disposal, we will employ the method explained at the end of section \ref{sec:ii} to suitably adjust the vacuum in order to eliminate the oscillations that our approximations spuriously generate and that amplify the power.

\subsection{Power spectrum for short-lived inflation in GR}

Let us first consider the evolution of primordial perturbations in a \emph{purely} classical relativistic cosmology with a short-lived single-field inflationary period. More specifically, we want to study the primordial power spectrum for the Mukhanov-Sasaki and tensor perturbations that propagate on a background that experiences inflation and presents a kinetically dominated pre-inflationary period. This background satisfies the Friedmann equation in the interval $(\eta_0,\eta_{\mathrm{end}}]$, where $\eta_0$ is any instant of time arbitrarily close to the Big-Bang singularity in this classical model. With this purpose, we approximate the dynamics of the gauge invariant perturbations, describing it by means of eq. \eqref{modeqgr}, which models the evolution on a kinetically dominated background that instantaneously transitions to a de Sitter phase.

If we decided to fix the mode solutions in the de Sitter phase $[\eta_i,\eta_{\mathrm{end}}]$ by using eqs. \eqref{muh} and \eqref{asymph}, we would simply arrive at the Bunch-Davies vacuum and a scale-invariant power spectrum. This choice of state would not encode any information from the pre-inflationary period, regardless of how long inflation lasts. This seems unphysical in those cases in which there exist observable Fourier scales $k$ of the order of the time-dependent mass in the pre-inflationary epoch. Therefore, for background solutions with short-lived inflation, this choice of a vacuum is not natural anymore.

Another possibility, motivated by our criterion for the selection of vacuum state, is to employ formulas \eqref{muh} and \eqref{asymph} instead in the kinetic epoch $(\eta_0,\eta_i)$ to try and fix the mode solutions there. As shown in Appendix \ref{app1}, this procedure leads to solutions of the form \eqref{mukin} with integration constants given by $C_k=0$ and $D_k=1$. These solutions have been previously selected to study the possible effects of short-lived inflation on the power spectrum, motivated by certain adiabatic considerations \cite{CPKL}. Actually, they can be obtained by demanding that the mode solutions correspond to the Poincar\'e vacuum in the limit $\eta\rightarrow\infty$. However, we should remark that the motivation and range of applicability of the formulas used here to select those solutions are much more general.

\begin{figure}
\centering
\includegraphics[width=6.5 cm]{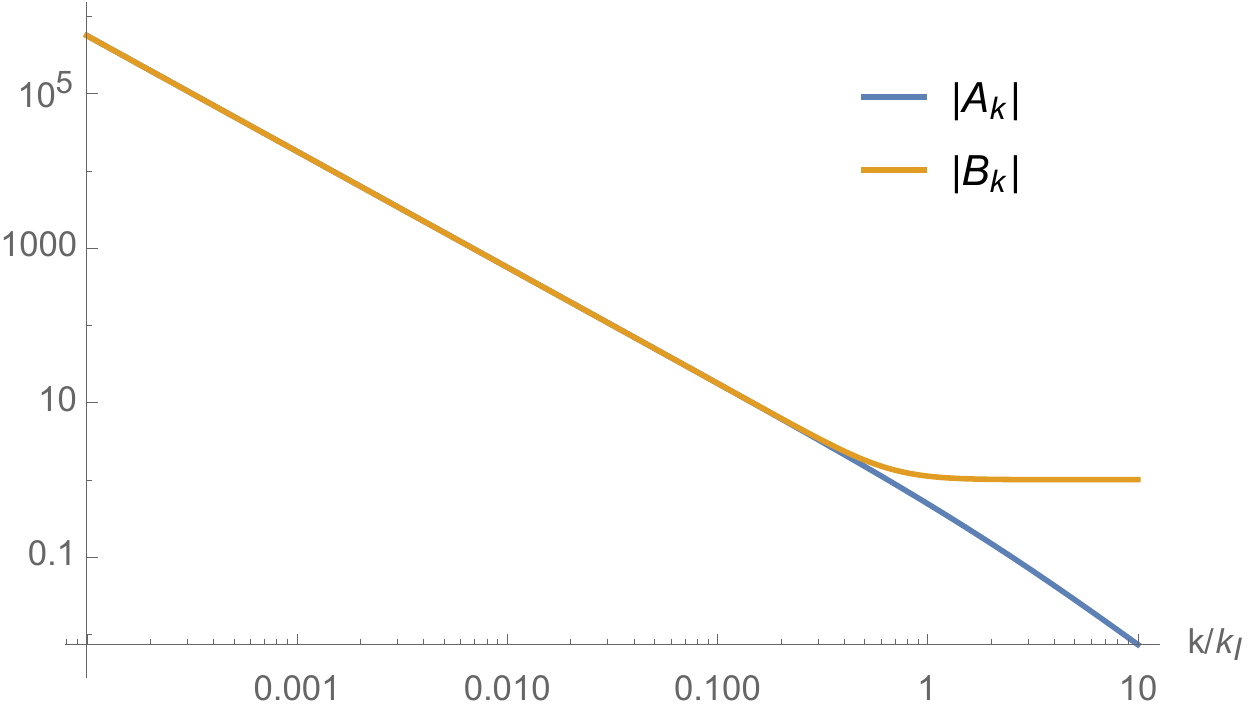}\includegraphics[width=9 cm]{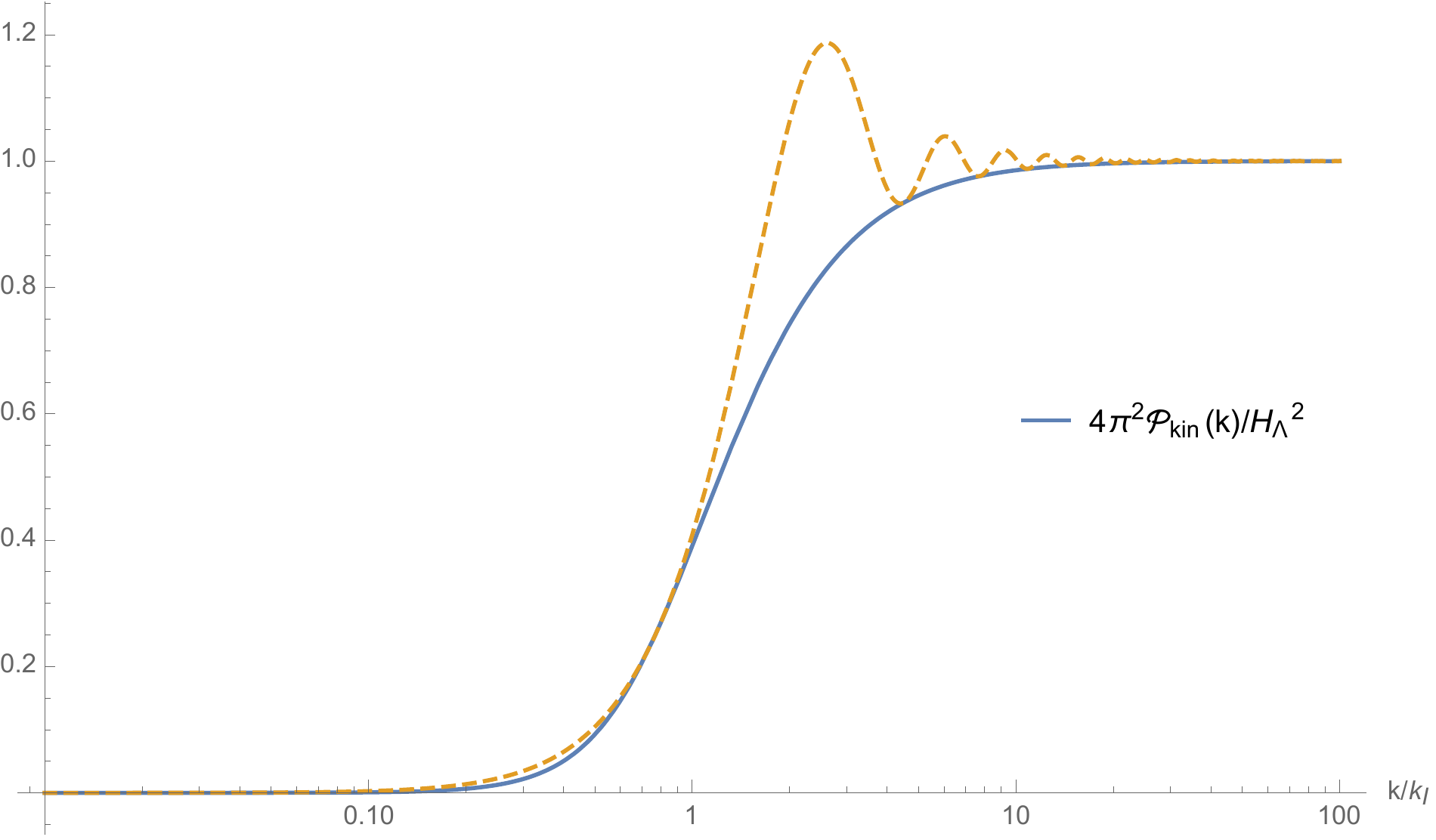}
\caption{{\emph{Left panel}}: Functions $|A_k|$ and $|B_k|$ of $k$ away from the ultraviolet regime $k\gg k_I$. {\emph{Right panel}}: Non-oscillating power spectrum $4\pi^2 \mathcal{P}_{\mathrm{kin}}(k)/H_{\Lambda}^2$ (continuous line), compared with its oscillating version corresponding to the choice $C_k=0$ and $D_k=1$ (dashed line).}
\label{fig2}
\end{figure}

Let us recall that the continuity of the solutions and of their first derivatives at $\eta_i$ leads to a unique primordial power spectrum [of the form \eqref{powerdesitter}]. For the considered choice of vacuum state, the constants $A_k$ and $B_k$ are those corresponding to the values $C_k=0$ and $D_k=1$ in eqs. \eqref{Ak} and \eqref{Bk}. If we define the Fourier scale $k_{I}=a_i H_\Lambda$, associated with the spacetime curvature effects at the onset of inflation, the resulting power spectrum is scale-invariant when $k\gg k_I$,  oscillates when $k\gtrsim 2k_I$, and presents a steep suppression for smaller $k$ (see the dashed line in the right panel of figure \ref{fig2}). We can understand this behavior by studying the different $k$-regions relative to $k_I$ separately.

In the region $k\gg k_I$, one can check that, for $C_k=0$ and $D_k=1$, 
\begin{equation}
A_k=\mathcal{O}(\sqrt{k_I/k}),\qquad B_k=e^{-3ik/(2k_{I})+i\pi/4}+\mathcal{O}(\sqrt{k_I/k}).
\end{equation}
We recall that the symbol $\mathcal{O}$ stands for terms of the order of its argument or subdominant. This result follows from the asymptotic behavior of the Hankel functions $H_{\nu}^{(j)}$ \cite{abram},
\begin{align}\label{Hankelasymp}
H_{\nu}^{(j)}(x)= \sqrt{\frac{2}{\pi x}}\exp{\left[-i(-1)^j (x-\nu\pi/2 -\pi/4)\right]}+\mathcal{O}(x^{-1}),\quad x\in\mathbb{R},\quad x\rightarrow\infty,
\end{align}
and it explains the scale-invariance of the power spectrum in the ultraviolet region (relative to $k_I$) at dominant order.

In the region $k\ll k_I$, we have that, for $C_k=0$ and $D_k=1$,
\begin{equation}
A_k=\sqrt{\frac{k\pi}{16k_I}}\bigg[\left(1+i\frac{k}{k_I}\right)H_0^{(2)}\left(\frac{k}{2k_I}\right)-\left(\frac{k_I}{k}+\frac{k}{k_I}\right)H_1^{(2)}\left(\frac{k}{2k_I}\right)\bigg]+A_k\mathcal{O}(k^2/k_I^2),
\end{equation}
\begin{equation}
B_k=\sqrt{\frac{k\pi}{16k_I}}\bigg[\left(1-i\frac{k}{k_I}\right)H_0^{(2)}\left(\frac{k}{2k_I}\right)-\left(\frac{k_I}{k}+\frac{k}{k_I}\right)H_1^{(2)}\left(\frac{k}{2k_I}\right)\bigg]+B_k\mathcal{O}(k^2/k_I^2).
\end{equation}
On the other hand, the following relations at dominant order (denoted with the symbol $\sim$) are valid for $|z|\ll 1$ \cite{abram}:
\begin{equation}
-iH_0^{(1)}(z)\sim iH_0^{(2)}(z)\sim \frac{2}{\pi}\ln{z},\qquad -iH_1^{(1)}(z)\sim iH_1^{(2)}(z)\sim -\frac{2}{\pi z}.
\end{equation}
Therefore, we get the dominant behavior
\begin{equation}
|B_k-A_k|\sim \sqrt{\frac{k^3}{\pi k^3_I}}\bigg|\ln{\left(\frac{k}{2k_I}\right)}\bigg|,
\end{equation}
which rapidly tends to zero as $k/k_I$ does. This explains the sharp suppression in the infrared region (relative to $k_I$).

Finally, in the intermediate region $k\gtrsim 2k_{I}$ one can check in eqs. \eqref{Ak} and \eqref{Bk} that, for $C_k=0$ and $D_k=1$, the relative phase between $A_k$ and $B_k$ contains the dominant term
$2k/k_I$. This ratio is of order one (or slightly bigger) in the considered region. It immediately follows that, unless $|A_k|$ and $|B_k|$ are oscillatory functions of $k$, the power spectrum \eqref{powerdesitter} oscilates for $k\gtrsim 2k_{I}$ owing to the commented phase. Indeed, in the left panel of figure \ref{fig2}  we plot $|A_k|$ and $|B_k|$ away from the ultraviolet region $k\gg k_I$ and we see that they do not oscillate. 

Looking at the dashed line in the right panel of figure \ref{fig2}, we also see that the oscillations in the intermediate region  $k\gtrsim 2k_{I}$ result in a mean amplification of the power in that sector. Taking into account that the amplitude of the positive-frequency solutions associated with this choice of vacuum state does not oscillate in the pre-inflationary epoch (see Appendix \ref{app2}), it seems reasonable to postulate that these oscillations are spurious and arise owing to the discontinuities that our approximations introduce in the instantaneous transition to a de Sitter phase. In this context, we refer to our discussion at the end of section \ref{sec:ii} and, accordingly, we proceed to eliminate this oscillatory behavior by considering instead the integration constants
\begin{equation}
A_k\rightarrow A_k^{\mathrm{kin}}=|A_k|,\qquad B_k\rightarrow B_k^{\mathrm{kin}}=|B_k|.
\end{equation}
With them, we obtain the non-oscillating power spectrum
\begin{equation}
\mathcal{P}_{\mathrm{kin}}(k)=\frac{H_{\Lambda}^2}{4\pi^2}\left(B^{\mathrm{kin}}_k-A^{\mathrm{kin}}_k\right)^2,
\end{equation}
that captures the non-amplified information about the oscillatory region of the original spectrum for $k\gtrsim 2k_{I}$, while respecting the infrared and ultraviolet behaviors. We show this non-oscillating spectrum in the right panel of figure \ref{fig2}, comparing it to its original version for the vacuum determined by the choice $C_k=0$ and $D_k=1$.

In summary, in a cosmological model with short-lived inflation ruled by the Friedmann equation of GR, we have singled out a set of mode solutions starting with a general criterion applied in the kinetically dominated regime, and then eliminating the oscillations that amplify the resulting power spectrum. Adopting this procedure, the effects on the primordial power spectrum of a purely classical pre-inflationary period in a short-lived inflationary cosmology are as follows. Given the scale $k_I$, equal to the inverse of the radius of the comoving Hubble horizon at the onset of inflation, one distinguishes two sectors of Fourier scales $k$ with very different properties. On the one hand, for $k\gg k_I$ the power spectrum is nearly scale-invariant, something that can be interpreted as the consequence that the perturbations are in the Bunch-Davies state at the end of inflation for those scales. On the other hand, as $k$ becomes comparable to $k_I$ and gets smaller,  the power spectrum gets highly suppressed. This indicates that, for such scales, the perturbations are in a(n excited) state different than the Bunch-Davies solution. 

Remarkably, this departure from the Bunch-Davies state only for Fourier scales smaller than $k_I$ happens even when {\emph{all the scales}} end up crossing the Hubble horizon as one approaches the Big-Bang singularity in the kinetically dominated phase. It is often argued that this crossing can excite the corresponding perturbations with respect to their Bunch-Davies state. Nonetheless, our results show that there exists in fact a natural mechanism in the kinetic epoch that can de-excite all modes with $k \gg k_I$. This means that, strictly speaking, the existence of an upper bound in the cosmological curvature (as e.g. in the case of effective LQC \cite{AGvacio2,AshNe}) is not completely necessary to explain a power suppression affecting only the infrared sector, as indicated by observations. Therefore, one needs additional features and details of the spectrum to distinguish effects in our primordial Universe that can be attributed exclusively to LQC.

\subsection{Power spectrum for hybrid LQC}

Let us finally analyze the primordial power spectrum of the cosmological perturbations in effective scenarios of hybrid LQC with short-lived inflation, using the same criterion for the choice of vacuum state as in the GR case. As we have motivated in section \ref{sec:iii}, we approximate the corresponding evolution of the Mukhanov-Sasaki and tensor perturbations from the bounce to the end of inflation by means of eq. \eqref{modeqlqc}. It contains a period $[\eta_B,\eta_0]$ in the vicinity of the bounce where the mass behaves as a P\"oschl-Teller potential, and then a classical kinetically dominated regime $(\eta_0,\eta_i)$, followed by a de Sitter inflation during the interval $[\eta_i,\eta_{\mathrm{end}}]$. Recalling the properties of the LQC backgrounds of interest, our approximation requires that the energy density at the instant $\eta_i$ be in the range $[10^{-12},10^{-9}]$. Taking, e.g., a value of the order of $10^{-9}$, i.e. the largest possible one, this requirement implies that $k_I=a_i H_{\Lambda}$ must be of the order of $10^{-3}$.

Formulas \eqref{muh} and \eqref{asymph}, associated with our preferred criterion for the choice of vacuum state, can be separatedly applied in any of the intervals $[\eta_B,\eta_0]$, $(\eta_0,\eta_i)$, or $[\eta_i,\eta_{\mathrm{end}}]$. The resulting set of mode solutions, that are continuous and have a continuous derivative at the matching points, changes with the interval chosen among the three commented options. As mentioned above, if we use formulas \eqref{muh} and \eqref{asymph} in the interval $[\eta_i,\eta_{\mathrm{end}}]$, we fix the Bunch-Davies vacuum, which is not the most natural choice if the pre-inflationary dynamics has affected the perturbations with observable scales $k$. On the other hand, if we instead focus our analysis on the classical kinetic period $(\eta_0,\eta_i)$, we obtain the state and power spectrum studied in the previous subsection in the context of a purely classical background. Therefore, this spectrum only encodes information from the effects that the pre-inflationary relativistic evolution of the background can have imprinted on the perturbations. In the context of effective LQC, quantum geometry phenomena can importantly affect the evolution of the perturbations near the bounce. One then reasonably expects that there should exist a different, yet physically preferred, vacuum state with a power spectrum that captures such LQC effects. This leads us to consider the last remaining possibility, namely, that we apply the considered formulas \eqref{muh} and \eqref{asymph} in the interval $[\eta_B,\eta_0]$ in order to fix the mode solutions for the perturbations.

As we show in Appendix \ref{app1}, the particular solution $\mu_k$ of eq. \eqref{modeqlqc} obtained with eqs. \eqref{muh} and \eqref{asymph} in $[\eta_B,\eta_0]$ corresponds to
\begin{equation}\label{hPTNO}
h_k=-i\alpha\tilde{k}-2\alpha x(1-x)	\frac{cd}{1+i\tilde{k}}\frac{{}_2 F_1\left(c+1,d+1;2+i\tilde{k};x\right)}{{}_2 F_1\left(c,d;1+i\tilde{k};x\right)},
\end{equation}
where $\tilde{k}=k/\alpha$ and we have introduced the following $k$-independent constants:
\begin{equation}\label{cd}
c=\frac{1}{2}\left(1+\sqrt{1+\frac{32\pi\rho_c}{3\alpha^2}}\right),\qquad d=\frac{1}{2}\left(1-\sqrt{1+\frac{32\pi\rho_c}{3\alpha^2}}\right).
\end{equation}
We recall that $\alpha$ and $x$ are defined in eqs. \eqref{alphau} and \eqref{xtime}. It is possible to show that the resulting normalized solution $\mu_k$ is that given by the integration constants $M_k=1/\sqrt{2k}$ and $N_k=0$ (up to a constant phase) in eq. \eqref{mupt}. Actually, this solution has been previously considered in the study of the primordial power spectrum arising from hybrid LQC, motivated by the imposition of adiabatic conditions in the far past region of the P\"oschl-Teller term \cite{wang,wang2}. These previous analyses adopted much simpler approximations for the kinetic epoch than those presented here. In addition, we emphasize that our criterion for the selection of vacuum state is supported by fundamental arguments applicable in more general scenarios than the presence of a time-dependent mass of P\"oschl-Teller form, and do not rest on particular asymptotic regimes of this mass (or of the associated cosmological background).

Starting with the normalized solutions $\mu_k$ in the interval $[\eta_B,\eta_0]$ for each $\vec{k}$, obtained by substituting in eq. \eqref{muh} the function $h_k$ of eq. \eqref{hPTNO}, we fix a set of mode solutions for the perturbations on the whole of our approximated LQC cosmology after the bounce, provided that we impose continuity up to the first time derivative. The resulting power spectrum is given by eq. \eqref{powerdesitter}, where the integration constants $A_k$ and $B_k$ are those specified in eqs. \eqref{Ak} and \eqref{Bk} in terms of $C_k$ and $D_k$, which are in turn fixed by means of eqs. \eqref{Ck} and \eqref{Dk}. In particular, we notice that
\begin{equation}
\mu_k(\eta_0)=\sqrt{-\frac{1}{2\text{Im}(h_k)(\eta_0)}},\qquad \mu_k'(\eta_0)=-h^{*}_k(\eta_0)\mu_k(\eta_0).
\end{equation}
From our expressions, one can realize that the behavior of the resulting spectrum depends on the considered region of $k$ relative to the scale $k_I\sim 10^{-3}$ at the onset of inflation, on the one hand, and to the quantities $a_0 H_0$ and $\alpha$, on the other hand. For transition times between the LQC and the classical regimes such that our approximations to the mass are reasonably good (i.e., around $t_0=0.4$), we have that the parameters $a_0 H_0$ and $\alpha$ are not far from one and three in Planck units, respectively. Remarkably, these values are of the same order as the time-dependent mass $s_h^{\mathrm{kin}}$ at the bounce, in hybrid LQC. In view of this, we define (e.g.) the scale $k_{\text{LQC}}=\alpha$ associated with the potentially observable LQC effects on the evolution of the perturbations near the bounce.

Let us first consider the sector of Fourier scales with $k\gg 1$. This covers the ultraviolet region, relative to both $k_I$ and $k_{\text{LQC}}$. Taking into account the asymptotic behavior of the hypergeometric function ${}_2 F_1 (a_1,a_2;a_3,z)$ for large $a_3$ \cite{abram}, we have
\begin{equation}
\text{Im}(h_k)=-k+\mathcal{O}(k^{-1}),\qquad \text{Re}(h_k)=\mathcal{O}(k^{-2}),
\end{equation}
and therefore
\begin{equation}
\mu_k(\eta_0)=\sqrt{\frac{1}{2k}}+\mathcal{O}(k^{-5/2}),\qquad \mu_k(\eta_0)=-i\sqrt{\frac{k}{2}}+\mathcal{O}(k^{-3/2}).
\end{equation}
Inserting this in eqs. \eqref{Ck} and \eqref{Dk} (which give the integration constants $C_k$ and $D_k$) and using the asymptotic behavior of the Hankel functions, we get
\begin{equation}
C_k=\mathcal{O}(k^{-1/2}),\qquad D_k=e^{ik/(2H_0a_0)-i\pi/4}+\mathcal{O}(k^{-1}),
\end{equation}
from which we obtain the following ultraviolet behavior for $A_k$ and $B_k$:
\begin{equation}
|A_k|=\mathcal{O}(k^{-1/2}),\qquad |B_k|=1+\mathcal{O}(k^{-1/2}).
\end{equation}
The resulting power spectrum is therefore scale-invariant for $k\gg 1$, in agreement with the standard result derived by using a Bunch-Davies state.

\begin{figure}
\centering
\includegraphics[width=7.5 cm]{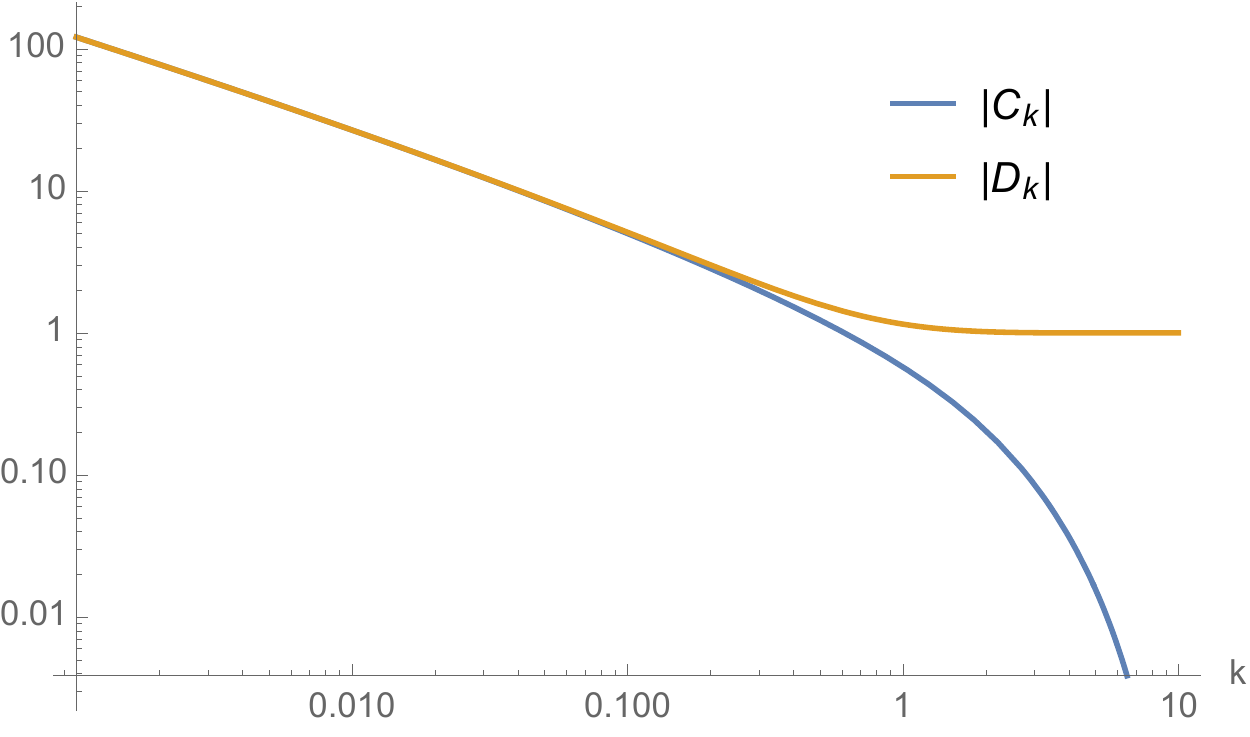}\includegraphics[width=7.5 cm]{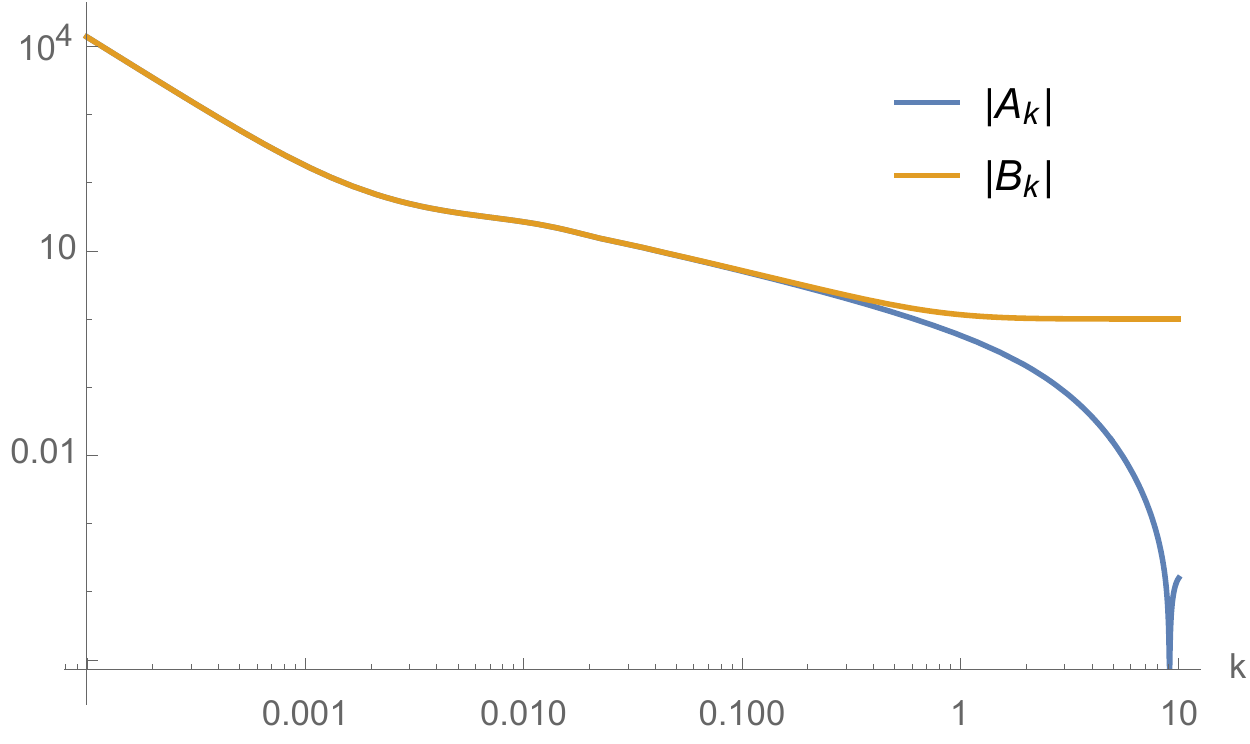}
\caption{The functions in both panels are computed with a transition time between the LQC and classical epochs equal to $t_0=0.4$. {\emph{Left panel}}: Functions $|C_k|$ and $|D_k|$ of $k$ away from the ultraviolet region $k\gg 1$. {\emph{Right panel}}: Functions $|A_k|$ and $|B_k|$ of $k$ away from the ultraviolet region $k\gg 1$.}
\label{fig3} 
\end{figure}

We next consider the region with $k\lesssim k_{\text{LQC}}$. Recalling the orders of magnitude that the scales $k_{\text{LQC}}$ and $k_I$ take in our system, in this region $3\gtrsim k\gg k_I$. 
It then follows that
\begin{equation}\label{Akasymp}
A_k=e^{3ik/(2k_{I})-i\pi/4}C_k \left[1+\mathcal{O}(\sqrt{k_I/k})\right]+D_k \mathcal{O}(\sqrt{k_I/k}),
\end{equation}
\begin{equation}\label{Bkasymp}
B_k=e^{-3ik/(2k_{I})+i\pi/4}D_k \left[1+\mathcal{O}(\sqrt{k_I/k})\right]+C_k \mathcal{O}(\sqrt{k_I/k}).
\end{equation}
To improve this asymptotic formula, we need more information about the relative order of magnitude of the integration constants $C_k$ and $D_k$. In the considered region, a priori, we do not expect this relative order to depend on $k/k_I$, because the formulas that determine those constants do not involve the scale $k_I$. We confirm our expectations in the left panel of figure \ref{fig3}, where we plot the values of $|C_k|$ and $|D_k|$ given by their analytical expressions.
Therefore, in the considered region $k\lesssim k_{\text{LQC}}$ and at dominant order with respect to $k/k_I$, we obtain the following behavior for the integration constants $A_k$ and $B_k$: 
\begin{equation}
A_k\sim e^{3ik/(2k_I)-i\pi/4}C_k,\qquad B_k\sim e^{-3ik/(2k_I)+i\pi/4}D_k.
\end{equation}
Notice the relative phase between them, that necessarily contains $3k/k_I$. Consequently, we expect the appearance of rapid oscillations in the resulting power spectrum around $k_{\mathrm{LQC}}\gg k_I$. Furthermore, we would also find slower, superposed oscillations if a $k$-dependent relative phase happened to exist between the constants $C_k$ and $D_k$.

\begin{figure}
\centering
\includegraphics[width=7.5 cm]{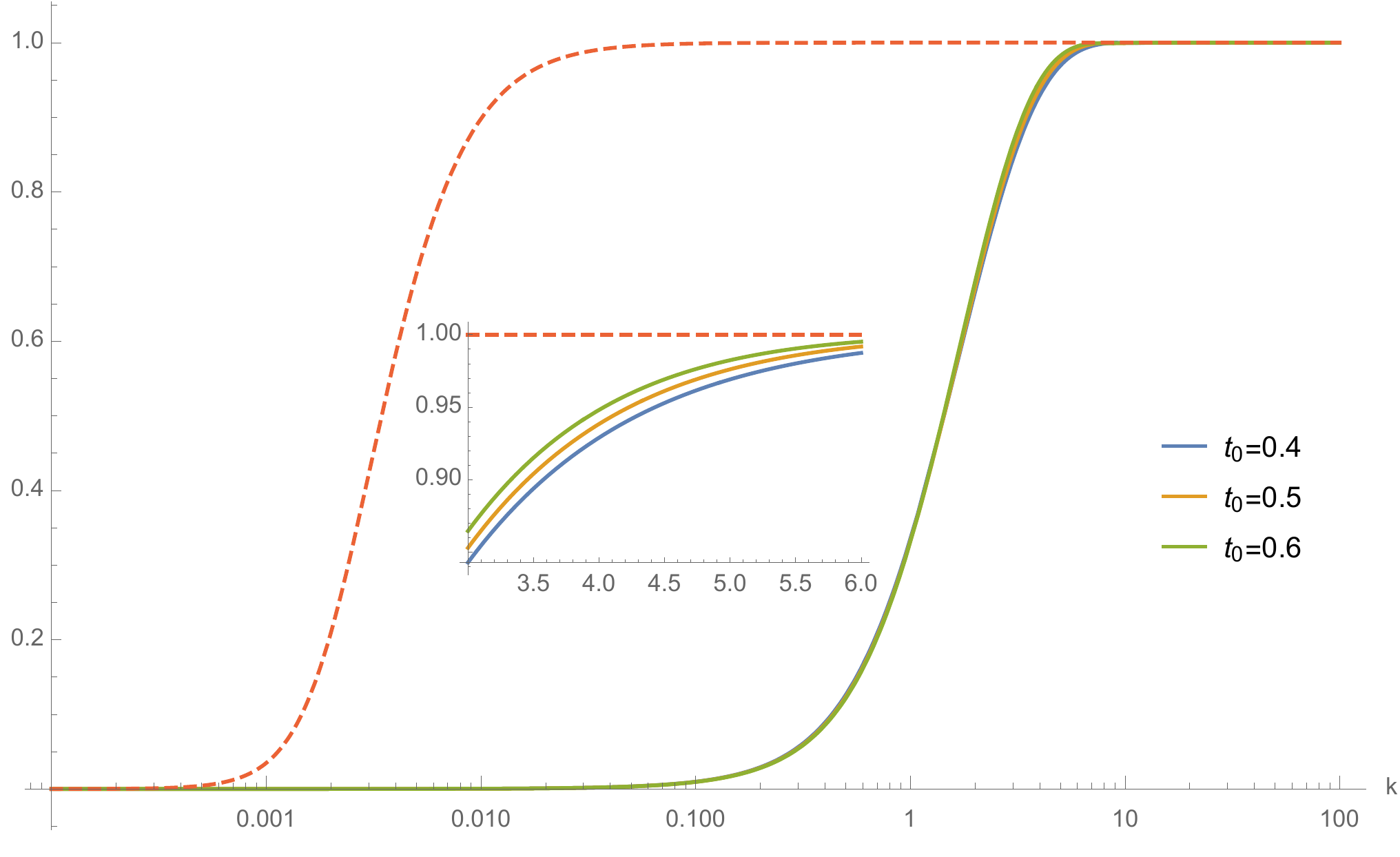}\includegraphics[width=7.5 cm]{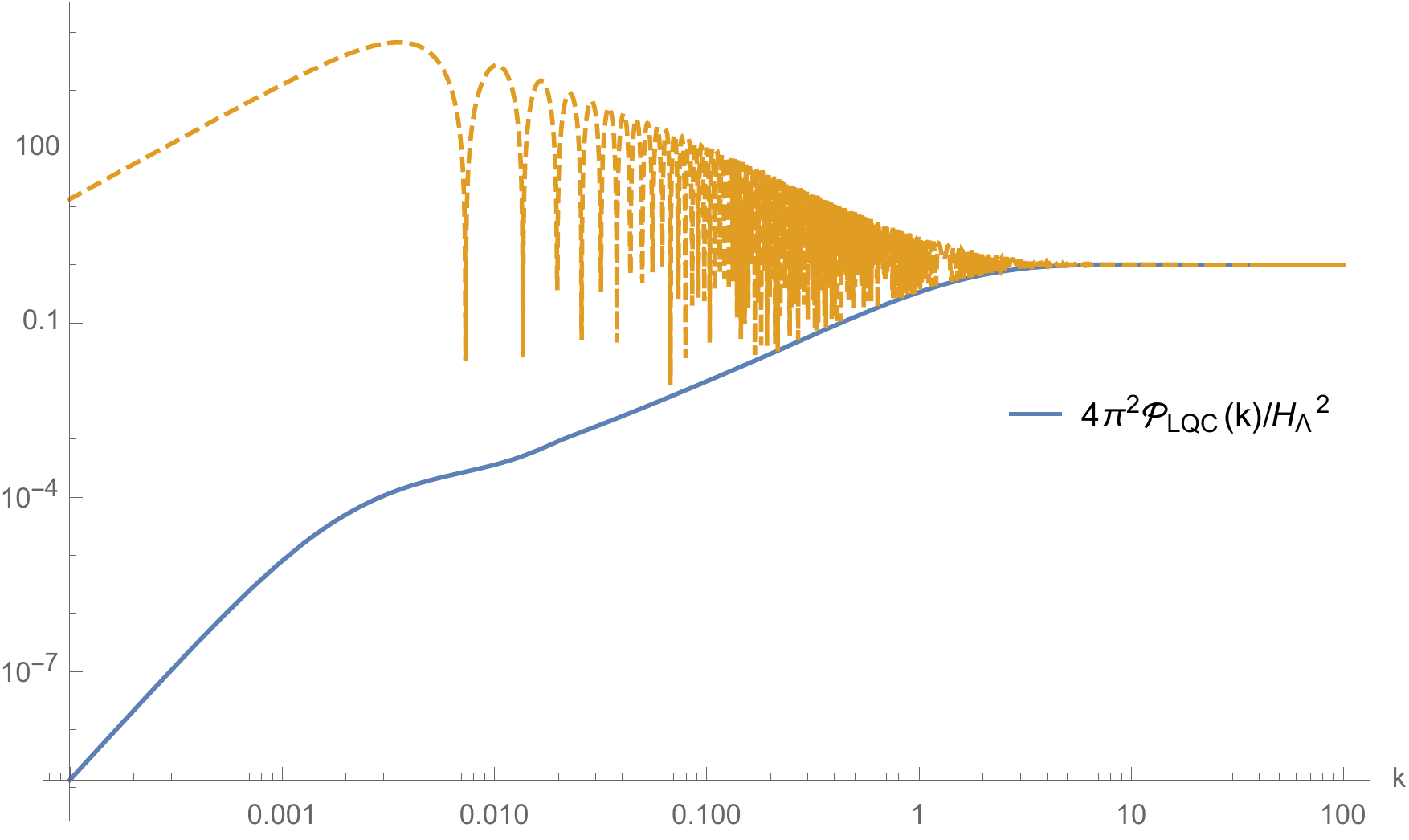}
\caption{{\emph{Left panel}}: Non-oscillating power spectrum $4\pi^2 \mathcal{P}_{\mathrm{LQC}}(k)/H_{\Lambda}^2$ in hybrid LQC  (continuous line), compared with its analog $4\pi^2 \mathcal{P}_{\mathrm{kin}}(k)/H_{\Lambda}^2$ in the GR context (dashed line). The inset shows the sensitivity on the choice of transition time $t_0$ between the LQC and classical epochs. {\emph{Right panel}}: Non-oscillating power spectrum $4\pi^2 \mathcal{P}_{\mathrm{LQC}}(k)/H_{\Lambda}^2$ for $t_0=0.4$ (continuous line), compared with its preliminary oscillating version before adjusting the dephasing between the integration constants $A_k$ and $B_k$ (dashed line). The difference between  the non-oscillatory spectrum and the oscillatory one at the minima of the latter is only apparent, owing to the limited resolution of the plot.}
\label{fig4} 
\end{figure}

The oscillations in the power spectrum suggest that we should still improve our choice of vacuum state, in the same spirit that was explained at the end of section \ref{sec:ii}. Indeed, our choice so far determines constants $A_k$ and $B_k$ that have non-oscillatory norms. Apart from the previous discussion, valid for $k$ around $k_{\text{LQC}}$, we show this in the right panel of figure \ref{fig3} for all scales that do not belong to the ultraviolet region $k\gg 1$. Furthermore, one can check (see Appendix \ref{app2}) that the amplitude of the positive-frequency solutions associated with this state does not oscillate in the vicinity of the bounce. So we expect that these oscillations are artificial and caused by the discontinuities that our approximations introduce in the change to the kinetically dominated and de Sitter classical epochs. In view of this, we follow {\emph{exactly the same procedure}} as in the case of the classical GR cosmology regime studied above, and redefine the integration constants $A_k$ and $B_k$ in such a way that the oscillations that amplify the resulting power spectrum are eliminated, while respecting its scale-invariant ultraviolet behavior. As we have already commented, the physical properties of the power spectrum crucially depend on this procedure that removes what we believe to be spurious oscillations. Explicitly, we introduce the following Bogoliubov transformation:
\begin{equation}
A_k\rightarrow A_k^{\mathrm{LQC}}=|A_k|,\qquad B_k\rightarrow B_k^{\mathrm{LQC}}=|B_k|,
\end{equation}
and consider the associated power spectrum
\begin{equation}
\mathcal{P}_{\mathrm{LQC}}(k)=\frac{H_{\Lambda}^2}{4\pi^2}\left(B^{\mathrm{LQC}}_k-A^{\mathrm{LQC}}_k\right)^2.
\end{equation}
We plot this non-oscillating spectrum $\mathcal{P}_{\mathrm{LQC}}$ in the left panel of figure \ref{fig4}. There, we also compare it with the spectrum $\mathcal{P}_{\mathrm{kin}}$ that we have obtained in the previous subsection for the case of a purely classical background, particularizing it to $k_I=2.8 \times 10^{-3}$. Recalling that $k_I=a_i H_{\Lambda}$, this choice leads (according to our criteria) to a preferred power spectrum, $\mathcal{P}_{\mathrm{kin}}$, for perturbations that propagate on a classical cosmology where inflation starts, roughly speaking, at the largest energy density that is possible among the effective LQC solutions of interest. In addition, in the case of $\mathcal{P}_{\mathrm{LQC}}$, we show the result for three different choices of the transition time $t_0$ between the LQC and classical epochs. The behavior of the resulting spectra is only slightly sensitive to this choice, for scales $k$ around $k_{\mathrm{LQC}}$. More importantly, we see that for $k< 2k_{\mathrm{LQC}}$, the primordial power $\mathcal{P}_{\mathrm{LQC}}$ gets strongly suppressed, in sharp contrast with the scale-invariant behavior of $\mathcal{P}_{\mathrm{kin}}$, that is only lost when $k<2k_{I}$. 

For completeness, in the right panel of figure \ref{fig4} we also show the oscillating power spectrum that would be obtained if we used the constants $A_k$ and $B_k$ resulting from our discontinuous matching between different cosmological epochs. In that panel we also compare this power spectrum with its non-oscillating version $\mathcal{P}_{\mathrm{LQC}}$. As we  anticipated in section \ref{sec:ii}, and confirmed in the pure relativistic scenario, the oscillations (which we view as artificial according to our hypotheses) produce a mean enhancement of the power of the perturbations at the end of inflation.

The plots in figure \ref{fig4} are restricted to the interval $k\in[10^{-4},10^2]$. For the considered effective LQC solutions with phenomenological application, this interval covers (with some margin in the infrared region and except for the most ultraviolet part in some cases) the window of wavenumbers that are directly observable nowadays \cite{Ivan,AGvacio2,no}. Roughly speaking, this window corresponds to $10^{-4} \mathrm{Mpc}^{-1} \lesssim k/a_\mathrm{today}\lesssim 10^{-1}\mathrm{Mpc}^{-1}$, where $a_\mathrm{today}$ is the current value of the scale factor \cite{planck}. Actually, the lower end of the observable interval of $k$ depends on the specific LQC solution that one selects. In particular, for those solutions which accumulate more e-folds from the bounce till the present, this end can be relatively large and range in $[10^{-1},1]$ \cite{Ivan,AGvacio2}, hence excluding the scale of inflation $k_I$ from the observable interval.

To our degree of approximation, the region where we have found power suppression, starting at distinct values of $k$, is the most important difference between the power spectra obtained in effective LQC and in purely classical backgrounds with short-lived inflation. It depends strongly on the difference between the scales $k_{\mathrm{LQC}}$ and $k_{I}$, which are related to the spacetime curvature around the vicinity of the bounce where LQC corrections are important and to the onset of inflation in relativistic cosmology, respectively. In fact, the order of magnitude of both of these scales is fixed for background solutions in effective LQC such that the largest sector of observable angular scales in the CMB may have been affected by quantum geometry effects. The suppression for $k<k_{\mathrm{LQC}}$ can be interpreted as a consequence of the vacuum excitation of these scales with respect to the Bunch-Davies state, owing to the presence of the bounce. Furthermore, the later period of deccelerated expansion does not de-excite these scales. 

It may seem surprising that such a narrow period of $0.4$ Planck times after the bounce, where LQC effects are important, can lead to such a great difference in the power spectrum when one compares it with the result obtained for pure GR with fast-roll inflation. However, there are two arguments that suggest this remarkable departure from a physical perspective. First of all, the evolution of the comoving Hubble parameter in the high-curvature regime is drastically different in the considered LQC solutions and in GR. In the former case, it vanishes at the bounce and then grows rapidly to reach a local maximum (of order one in Planck units), after which the classical kinetically dominated epoch starts. In the pure GR model with fast-roll inflation, on the other hand, the comoving Hubble parameter grows unboundedly to the past and displays a diverging behavior as one approaches the Big-Bang singularity. Therefore, in the high-curvature regime, the perturbations with physical wavelengths of the same order or slightly larger than the Planck unit, which are those with frequencies that can best feel the quantum-induced variations of the curvature in this regime, cross the Hubble horizon twice in the LQC scenario. This happens regardless of how narrow the interval with strong LQC effects is. On the contrary, this crossing only happens once in the analog high-curvature regime of the considered GR model. Secondly, the type of LQC solutions for the background that we are considering all display a local minimum of the comoving Hubble parameter, roughly at the onset of inflation, that is at least $10^{3}$ smaller than its maximum right after the bounce. Since at those times the evolution of the LQC background is indistinguishable from the relativistic one, its analog GR model (for which our comparison is meaningful) shares this property. The great difference between the values of the local maximum of the comoving Hubble parameter, present only in LQC, and its minimum at the onset of inflation, present also in GR, is precisely translated into the difference between the scales $k_{\mathrm{LQC}}$ and $k_{I}$, that are the ones that affect the evolution of the perturbations and leave their imprint in the suppressed power spectra in both models (for our choice of vacua). It is quite remarkable that this great difference is a genuine signature of the background solutions of LQC with phenomenological interest, namely those for which inflation is just long enough so that the most infrared scales that are observable nowadays might have been affected by quantum geometry phenomena around the bounce.

Let us remark that the power spectra $\mathcal{P}_{\mathrm{LQC}}$ and $\mathcal{P}_{\mathrm{kin}}$ have been obtained using the same procedure for the choice of vacuum state, even though this leads to different vacua in each of the considered cosmological models (as it should be the case, since the dynamical behavior of the relativistic background is drastically different from the LQC one in the high-curvature regime). That is, starting with eqs. \eqref{asymph} and \eqref{recursion} applied in the earliest period of the cosmological evolution, we have fixed the respective vacua by selecting integration constants $(A_k^{\mathrm{LQC}},B^{\mathrm{LQC}}_k)$ and $(A^{\mathrm{kin}}_k,B^{\mathrm{kin}}_k)$ that avoid, in the same specific way, the oscillations that amplify the resulting power spectrum. In this sense, the differences between both power spectra shown in our analysis reflect genuine traces of the loop quantum bounce, distinguishing them from classical pre-inflationary effects. 

Finally, it is worth noticing that our results would have been very different had we not followed the criteria put forward in subsection \ref{sec:iv1} to characterize the vacuum, motivated by our theoretical considerations. The redefinition that we have made of the constants $A_k$ and $B_k$ is just devised to capture the main properties of this original vacuum that are not due to the discontinuities of our analytical approximation. Thus, our results depend crucially and non-trivially on the criterion proposed to select the vacuum state. In fact, from all our previous discussions, it follows that this vacuum is able to capture the main features that explain the notable differences in the power spectrum obtained for GR and LQC even before our redefinition process. Namely, in the power spectrum derived from the constants $A_k$ and $B_k$, scale invariance is broken (even if in an oscillatory way) starting from one of the two aforementioned scales: the curvature at the onset of inflation in the case of pure GR, and the curvature at the bounce in the case of LQC. Moreover, the power suppression that we have finally found in both cases is not a property that would be generally shared by other original choices of vacuum. This is so even if, for such alternative vacua, one eliminated as well the oscillations that may be generated by our approximations along the procedure presented in section \ref{sec:ii}.

\section{Discussion}\label{sec:v}

This work is the first analytic comparison between the main effects on the primordial power spectrum arising from a single-field inflationary cosmology with a fast-roll period within GR, on the one hand, and from the cosmological evolution of effective LQC backgrounds of phenomenological interest, on the other hand. In this last case, we have focused our full attention on the dynamical equations of the Mukhanov-Sasaki and tensor perturbations that result from the hybrid loop quantization approach. These equations have the same structure as in GR, but with a different and very specific time-dependent mass. Both of the considered cosmological backgrounds (i.e., the classical relativistic and the effective LQC backgrounds) for the propagation of the perturbations coincide a few Planck times after the instant that would correspond in effective LQC to the bounce. They display a classical kinetically dominated era that transitions into a slow-roll inflation that is short-lived. 

In order to study the problem analytically, as a first approximation, we have ignored any transition effects between the kinetic and slow-roll regimes, and described the classical cosmological evolution as a Friedmann solution with a massless scalar field that is instantaneously matched with a de Sitter spacetime. Furthermore, in the LQC case we have approximated the time-dependent mass function for the dynamics of the perturbations in the vicinity of the bounce by some P\"oschl-Teller potential, selected to improve similar approximations that have been previously proposed in the literature \cite{wang}. We have then derived general analytic formulas to characterize any choice of vacuum state for the perturbations and its resulting primordial power spectrum, for both of our considered classical and LQC cosmologies.

The form and properties of this power spectrum depend critically on the specific choice of vacuum state. It is then evident that a robust comparison between the effects of the different cosmological evolutions on the spectrum requires a common criterion for the choice of such a state. Inspired by recent studies in hybrid LQC \cite{msdiag,ms-corr}, and by the physical interest of attaining a non-(rapidly-)oscillating power spectrum \cite{no,NO-analy}, we have proposed a criterion that is applicable for both of our types of backgrounds (classical and in effective LQC). The starting point in both cases is to choose the state that results from imposing an asymptotic, time-dependent (non-instantaneous) diagonalization of the Hamiltonian during the earliest period of cosmological evolution in the respective approximate models. This procedure can be physically motivated from hybrid LQC, and its application only finds one obstruction for our background models, namely the discontinuities that our approximations introduce in the matching between the different epochs. These discontinuities might introduce some rapidly changing frequencies in the vacuum solutions, that have to adapt themselves fast to the discontinuities of the evolution.  Therefore, one should expect that the resulting power spectrum can present some undesired properties, such as the kind of unwanted oscillations that we try to avoid. Nonetheless, we have argued in favor of a well-motivated procedure to redefine the considered choice of vacuum so as to eliminate from its primordial spectrum the oscillations that amplify the power (in the sense explained in section \ref{sec:ii}), which happen to be easily identifiable in this situation. The power spectrum obtained in this way captures the essential properties of the originally chosen vacuum that do not get artificially altered by those oscillations.

The main differences that our analysis finds between both power spectra (namely, that arising from a purely classical GR background with fast-roll effects and that from effective LQC) are notorious. They depend on the difference between two natural scales of the cosmological models, $k_I$ and $k_{\mathrm{LQC}}$, that are respectively related to the spacetime curvature at the onset of inflation and at the bounce. With our choice of vacuum state for the perturbations, the power spectrum of the studied classical backgrounds does not oscillate in the Fourier scale $k$, and displays a drastic infrared suppression for $k<k_I$, while displaying scale-invariance for larger $k$. This scale-invariant behavior in the ultraviolet combined with a strong suppression in the infrared has already been reported in past investigations, e.g. in refs. \cite{CPKL,RS,Ramirez,DVS2,HHHL,SA}, however often using choices of vacuum state that lead to an artificially oscillating power for $k\gtrsim k_I$.  On the other hand, our non-oscillating spectrum corresponding to the effective LQC cosmologies (in the sector of phenomenological interest) has strongly suppressed power for $k<k_{\mathrm{LQC}}$ and is scale-invariant for larger $k$. Since all those cosmologies must involve a (physically significant) classical kinetic epoch prior to slow-roll inflation, it follows that $k_{\mathrm{LQC}}\gg k_I$ for any of them. Therefore, we conclude that the main effects of the loop quantum bounce on the primordial perturbations translate into a huge infrared suppression in their spectrum (viewable in practice as if there existed a cutoff) that covers a much wider region of scales than it would if the perturbations were solely affected by a kinetically dominated pre-inflationary dynamics within GR. In this sense, it is worth commenting that, for the set of effective LQC solutions that we have considered (as well as for their analogues within GR), the scale of inflation $k_I$ lies either in the very lower end of the window of directly observable scales $k$, or just out of it for those backgrounds with the largest number of e-folds of expansion. One would then expect that the net suppression on the angular power spectrum of temperature anisotropies should turn out to be much greater if one takes into account all of the LQC effects rather than only the classical fast-roll ones.

Although yet very approximate, it is enlightening to compare our analytic results with previous ones attained in the literature of hybrid LQC by solving numerically the exact equations for the cosmological background and perturbations \cite{no,hybr-pred}. There, the predicted power spectrum is also non-oscillating and displays a clear infrared suppression. However, this strong power suppresion starts at a Fourier scale that is of the typical order of $k_I\sim 10^{-3}$, rather than that of $k_{\mathrm{LQC}}$. In view of our conclusions, this behavior seems to indicate that the choice of vacuum state adopted in those works is only capable of capturing the classical fast-roll effects on the evolution of the perturbations, but not the LQC effects associated with a bounce in the high-curvature regime. Other remarkable investigations in LQC \cite{AshNe,AGvacio2} have also predicted an infrared suppression in the power spectrum, obtained after averaging its fast superimposed oscillations. In this case, the suppression does start from a Fourier scale that can be related to the spacetime curvature at the bounce. However, the theoretical framework underlying these works is rather different from the hybrid approach. It is usually called the dressed-metric formalism \cite{AAN3,AAN1,AAN2}, and the corresponding time-dependent mass in effective LQC is not the same. For the effective LQC solutions that we have considered, perhaps the most remarkable difference is the negativity of this mass at the bounce, compared to the positivity of $s_h^{(r)}$. It would be interesting to investigate if one can apply similar approximations and choice of vacuum state to analytically study the dressed-metric case, comparing the resulting power spectrum with the hybrid one.

Concerning our approximations in the high-curvature regime, we should note that these are yet preliminary in the case of hybrid LQC cosmology. In fact, the errors that they introduce on the time-dependent mass of the perturbations are, in the best case scenario, of a $15\%$ in the vicinity of the bounce. These errors, in turn, are expected to have non-negligible consequences on the choice of vacuum state. However, our approximations seem good enough so as to permit that the power spectrum truly reflects the effects of the loop quantum bounce, when compared to a pure GR scenario, namely to show the departure from scale invariance (in the form of a suppression of power, for a suitable choice of vacuum state) at a very different and much larger wavenumber scale than in the GR case. This claim is supported by figure \ref{fig4}, where the same qualitative behavior is found for different approximations to the time-dependent mass around the bounce (codified in different choices of $t_0$), for which the errors can grow beyond a $25\%$. It would be desirable to try and investigate in future work how to improve our analytical investigation of the evolution of the perturbations in the vicinity of the bounce, characterizing an appropriate vacuum state that is optimally adapted to the dynamics of the cosmological background in that region. Furthermore, these analytical studies should be complemented, for completeness and comparison, with a numerical evolution of the initial conditions selected by our vacuum proposal for the exact LQC background (obtained as well via numerical integration of the modified Friedmann equations).

Finally, is worth commenting on the fact that, because of the simplicity of our approximations, we have been forced to ignore in this work all relevant inflationary aspects related with quantities such as the spectral index and the tensor-to-scalar ratio, which reflect the departure of the slow-roll period from an exact de Sitter evolution. This mainly depends on the specific form of the inflaton potential. We believe that the analytical techniques presented in this work can be further developed to incorporate such inflationary aspects, taking into account the typical values of the slow-roll parameters in the effective LQC solutions of interest. This kind of studies would then provide a more realistic power spectrum for potential comparison with the CMB observations, while facilitating a clear understanding of the reasons underlying its behavior.

\appendix

\section{Interferences in the power spectrum produced by the matching}\label{app0}

Let us consider two consecutive intervals in the cosmological evolution, each of them associated with a different mass term for the primordial perturbations, mass that therefore may display some discontinuity at the conformal time $\eta_m$ where the two intervals meet. Furthermore, for $\eta<\eta_m$, let this mass term be a fixed function of time $s(\eta)$, while for $\eta\geq\eta_m$ we let it be any element of a one-parameter family $s_{\lambda}(\eta)$, with $\lambda\in\mathbb{R}^+$ and such that $s_{0}(\eta)=s(\eta)$. For the dynamics defined by each mass $s_{\lambda}(\eta)$, generic choices of positive-frequency solutions $\{\mu^{\lambda}_k\}$ for the perturbations can be expressed in terms of a given basis of normalized solutions $\{(\nu^{\lambda}_k)^{*}\}$ (and their complex conjugates) as:
\begin{align}
\mu_k^{\lambda}(\eta)=A_k^{\lambda}\nu_k^{\lambda}(\eta)+B_k^{\lambda}(\nu_k^{\lambda})^{*}(\eta),
\end{align}
where the complex constant coefficients are normalized so that $|B^{\lambda}_k|^2-|A^{\lambda}_k|^2=1$. With this notation, in the past of $\eta_m$ the positive-frequency solution is just $\mu_k^{0}$, whereas, in the future of this instant of time, the solution becomes the member of the family $\mu_k^{\lambda}$ corresponding to the mass $s_{\lambda}(\eta)$. We can construct a state for the perturbations through the whole cosmological evolution if we impose continuity of its positive-frequency solutions at $\eta_m$. This yields:
\begin{align}
A_k^{\lambda}=i\left[\nu_k^{0}(\nu_k^{\lambda})^{*\prime}-(\nu_k^{\lambda})^{*}\nu_k^{0\prime}\right]_{\eta_m}A^{0}_k+i\left[(\nu_k^{0})^{*}(\nu_k^{\lambda})^{*\prime}-(\nu_k^{\lambda})^{*}(\nu_k^{0})^{*\prime}\right]_{\eta_m}B^{0}_k ,
\end{align}
\begin{align}
B_k^{\lambda}=i\left[\nu_k^{\lambda}\nu_k^{0\prime}-\nu_k^{0}\nu_k^{\lambda\prime}\right]_{\eta_m}A^{0}_k+i\left[\nu_k^{\lambda}(\nu_k^{0})^{*\prime}-(\nu_k^{0})^{*}\nu_k^{\lambda\prime}\right]_{\eta_m}B^{0}_k,
\end{align}
where the subscripts after the square brackets indicate evaluation at $\eta_m$. Let us assume that the initial state, in the past of $\eta_m$, is chosen so that $A_k^0$ and $B_k^0$ have the same phase. It is clear then that, for generic $s_{\lambda}$, the coefficients $A_k^{\lambda}$ and $B_k^{\lambda}$ will generally present some dephasing unless $\lambda=0$, namely unless the mass function $s(\eta)$ is unchanged in the transition time. Since any definition of primordial power spectrum directly involves the squared norm of the positive-frequency solutions, if it is evaluated after the instant $\eta_m$ for $\lambda\neq 0$, then the discussed dephasing will be reflected in the spectrum as an interference term, possibly giving rise to $k$-dependent oscillations.

The above discussion clearly supports our arguments that an artificial (and to certain extent discontinuous) matching of cosmological epochs, performed in order to handle the problem of obtaining a family of positive-frequency solutions for the perturbations, translates into dephased constants of integration for these solutions in the latest of these epochs. Such a dephasing directly affects the (non-)oscillatory behavior of the primordial power spectrum. Consequently, in this work we choose to eliminate it at the end of our computations.

\section{Fixing a vacuum state with the asymptotic expansion of $h_k$}\label{app1}

Let us first show how we can fix the vacuum state associated with the choice of normalized mode solution \eqref{mukin} with $C_k=0$ and $D_k=1$ in the kinetically dominated interval $(\eta_0,\eta_{i})$, by applying eqs. \eqref{muh} and \eqref{asymph}. In this period, the function $h_k$ satisfies by construction
\begin{equation}
h_k'=k^2+\frac{1}{4y^2}+h_k^2,\qquad y=\eta-\eta_0+\frac{1}{2H_0 a_0}.
\end{equation}
If we introduce the change of function
\begin{equation}\label{hsigma}
h_k=-k\left[\frac{1}{2T_k}+\frac{d}{dT_k}(\log{\sigma})\right],\qquad T_k=ky,
\end{equation}
we are led to the Bessel equation
\begin{equation}\label{sigmaeq}
T_k^2\frac{d^2\sigma}{dT_k^2}+T_k\frac{d\sigma}{dT_k}+T_k^2\sigma=0.
\end{equation}
The general solution os this equation can be expressed as a linear combination of the Hankel functions of zeroth order $H_0^{(1)}(ky)$ and $H_0^{(2)}(ky)$. On the other hand, one can check that the asymptotic restriction imposed by eqs. \eqref{asymph} and \eqref{recursion} on $h_k$ implies
\begin{equation}
h_k=-ik\left[1+\mathcal{O}(k^{-2})\right],
\end{equation}
where the subdominant terms $\mathcal{O}(k^{-2})$ must form a series in inverse powers of $T_k=ky$. Comparing this condition on $h_k$ with eq. \eqref{hsigma}, 
and integrating the resulting equality in the asymptotic regime $k\rightarrow\infty$, we have that $\sigma$ must admit an asymptotic expansion of the form
\begin{equation}\label{sigmasymp}
\sigma=e^{iT_k}(T_k)^{-1/2}\sum_{n=0}^{\infty} (T_k)^{-n}\sigma_{n}
\end{equation}
when $T_k\rightarrow\infty$, where $\sigma_{n}$ are certain constants. Recalling the asymptotic behavior of the Hankel functions for large arguments, given in eq. \eqref{Hankelasymp}, we conclude that the only solution of eq. \eqref{sigmaeq} with the desired asymptotic behavior is
\begin{equation}
\sigma=\sigma_0 \sqrt{\frac{\pi}{2}}e^{i\pi/4}H_0^{(1)}(ky),
\end{equation}
for arbitrary constant $\sigma_0$. Therefore, the asymptotic eqs. \eqref{asymph} and \eqref{recursion} allow us to fix $h_k$ as
\begin{equation}
h_k=k\frac{H^{(1)}_1 (ky)}{H^{(1)}_0 (ky)}-\frac{1}{2y}.
\end{equation}
This selects a specific normalized solution $\mu_k$ of the dynamical eq. \eqref{skin} for the perturbations via eq. \eqref{muh}. Explicitly, one can check that the squared norm of this solution is
\begin{equation}
|\mu_k|^2=\frac{\pi y}{4}|H_0^{(2)}(ky)|^2.
\end{equation}
In the derivation of this formula, we have used the identity \cite{abram}
\begin{equation}\label{wronsk}
H_1^{(1)}(ky)H_0^{(2)}(ky)-H_0^{(1)}(ky)H_1^{(2)}(ky)=-\frac{4i}{\pi ky}.
\end{equation}
Finally, the phase of the mode solution can be determined in eq. \eqref{muh} noting that
\begin{equation}
\text{Im}(h_k)=-\text{arg}\left[H^{(1)}_0 (ky)\right]'=\text{arg}\left[H^{(2)}_0 (ky)\right]'.
\end{equation}
Thus, our asymptotic expansion for $h_k$ indeed fixes the vacuum state associated with the mode solutions \eqref{mukin} with $C_k=0$ and $D_k=1$, up to a constant irrelevant phase.

Let us now show that the asymptotic eqs. \eqref{asymph} and \eqref{recursion} fix as well the vacuum state associated with our normalized mode solutions in the interval $[\eta_B,\eta_0]$ near the bounce. We recall that these mode solutions are obtained by inserting in eq. \eqref{muh} the function $h_k$ given in eq. \eqref{hPTNO}. In this case, $h_k$ satisfies by construction that
\begin{equation}
h_{k}'=k^2+\frac{U_0}{\cosh^{2}{\alpha(\eta-\eta_B)}}+h_k^2,
\end{equation}
where $U_0$ and $\alpha$ were defined in eq. \eqref{alphau}. If we introduce the change of function
\begin{equation}\label{hu}
h_k=ik(1-2x)-2\alpha x(1-x)\frac{d}{dx}(\log{u_k}),
\end{equation}
with $x$ given in eq. \eqref{xtime}, then we obtain the hypergeometric equation
\begin{equation}\label{hypereq}
x(1-x)\frac{d^2 u_k}{dx^2}+\left[b_3^k-(b_1^k+b_2^k+1)x\right]\frac{du_k}{dx}-b_1^k b_2^k u_k=0.
\end{equation}
The parameters $b_1^k$, $b_2^k$, and $b_3^k$ are defined in general as
\begin{equation}
b_3^k=1-\frac{ik}{\alpha}
\end{equation}
\begin{equation}
b_1^k=\frac{1}{2}\left(1+\sqrt{1+\frac{4U_0}{\alpha^2}}\right)-\frac{ik}{\alpha},\qquad b_2^k=\frac{1}{2}\left(1-\sqrt{1+\frac{4U_0}{\alpha^2}}\right)-\frac{ik}{\alpha}.
\end{equation}
Clearly, they reproduce those given in eqs. \eqref{a3} and \eqref{a12} for $U_0=8\pi\rho_c/3$. The general solution of eq. \eqref{hypereq} is a linear combination of the two hypergeometric functions \cite{abram}
\begin{equation}
{}_2 F_1\left(b_1^k,b_2^k;b_3^k;x\right),\quad \quad x^{ik/\alpha}{}_2 F_1\left(b_1^k-b_3^k+1,b_2^k-b_3^k+1;2-b_3^k;x\right).
\end{equation}
In our case, we have
\begin{equation}
b_1^k-b_3^k+1=\frac{1}{2}\left(1+\sqrt{1+\frac{4U_0}{\alpha^2}}\right),\quad b_2^k-b_3^k+1=\frac{1}{2}\left(1-\sqrt{1+\frac{4U_0}{\alpha^2}}\right).	
\end{equation}
Again, for $U_0=8\pi\rho_c/3$ these formulas reproduce the respective values of the constants $c$ and $d$ in eq. \eqref{cd}. 

Taking into account that, in terms of $x$,
\begin{equation}
\frac{U_0}{\cosh^{2}{\alpha(\eta-\eta_B)}}=4U_0 x(1-x),
\end{equation}
it is easy to check that the asymptotic behavior required by eqs. \eqref{asymph} and \eqref{recursion} for $h_k$ implies 
\begin{equation}
h_k=-ik\left[1+\frac{2U_0}{k^2}x(1-x)+\mathcal{O}(k^{-3})\right].
\end{equation}
Furthermore, the subdominant terms $\mathcal{O}(k^{-3})$ must depend polynomically on $x$. If we introduce this behavior in eq. \eqref{hu} and integrate the result in the asymptotic regime of large $k$, we get that $u_k$ must admit an asymptotic expansion of the form
\begin{equation}\label{uasymp}
u_k=x^{ik/\alpha}\sum_{n=0}^{\infty}v^k_n x^n,	
\end{equation}
where the complex constants $v^k_n$ are linear combinations of inverse powers of $k$. At this stage, we recall that $x<1$ by the very construction of this variable, and that the hypergeometric functions that appear in the solutions of eq. \eqref{hypereq} admit representations as power series of $x$ that are convergent precisely when $x<1$ \cite{abram}. In view of this, we impose that the behavior \eqref{uasymp} holds for all $k$. The only solution of eq. \eqref{hypereq} that can depend on $x$ as in eq. \eqref{uasymp} is
\begin{equation}
u_k=v^k_0 x^{ik/\alpha}{}_2 F_1\left(b_1^k-b_3^k+1,b_2^k-b_3^k+1;2-b_3^k;x\right).
\end{equation}
Introducing this solution in eq. \eqref{hu}, we finally obtain
\begin{align}
h_k=-ik -2\alpha x(1-x)	\frac{(b_1^k-b_3^k+1)(b_2^k-b_3^k+1)}{2-b_3^k}\frac{{}_2 F_1\left(b_1^k-b_3^k+2,b_2^k-b_3^k+2;3-b_3^k;x\right)}{{}_2 F_1\left(b_1^k-b_3^k+1,b_2^k-b_3^k+1;2-b_3^k;x\right)},
\end{align}
which indeed coincides with eq. \eqref{hPTNO} when $U_0=8\pi\rho_c/3$.

\section{Non-oscillatory behavior of the vacuum in its earliest stage of evolution}\label{app2} 

In this appendix we show how the positive-frequency solutions that we have selected with the asymptotic expansion of $h_k$ display a non-oscillatory behavior in the epochs in which the asymptotic expansion applies, both in the case of pure GR with fast-roll inflation, for which
\begin{equation}\label{app2eq1}
\mu_k=\sqrt{\frac{\pi y}{4}} H^{(2)}_0 (ky), \qquad y=\eta-\eta_0+\frac{1}{2H_0 a_0},
\end{equation}
and in the case of the effective LQC solutions of phenomenological interest, where we have
\begin{equation}\label{app2eq2}
\mu_k=\sqrt{-\frac{1}{2\text{Im}(h_k)}}e^{i\int_{\eta_0}^\eta d\tilde\eta \text{Im}(h_k)(\tilde{\eta})},
\end{equation}
with $h_k$ given in eq. \eqref{hPTNO}. These epochs are the earliest stages of their evolution, which correspond respectively to the intervals $(\eta_0,\eta_i)$ in conformal time for GR, and $[0,0.4]$ in proper time for LQC. In this last case, we have already adopted the value $t_0=0.4$ that provides the best approximation to the time-dependent mass around the bounce.

\begin{figure}
\centering
\includegraphics[width=7.5 cm]{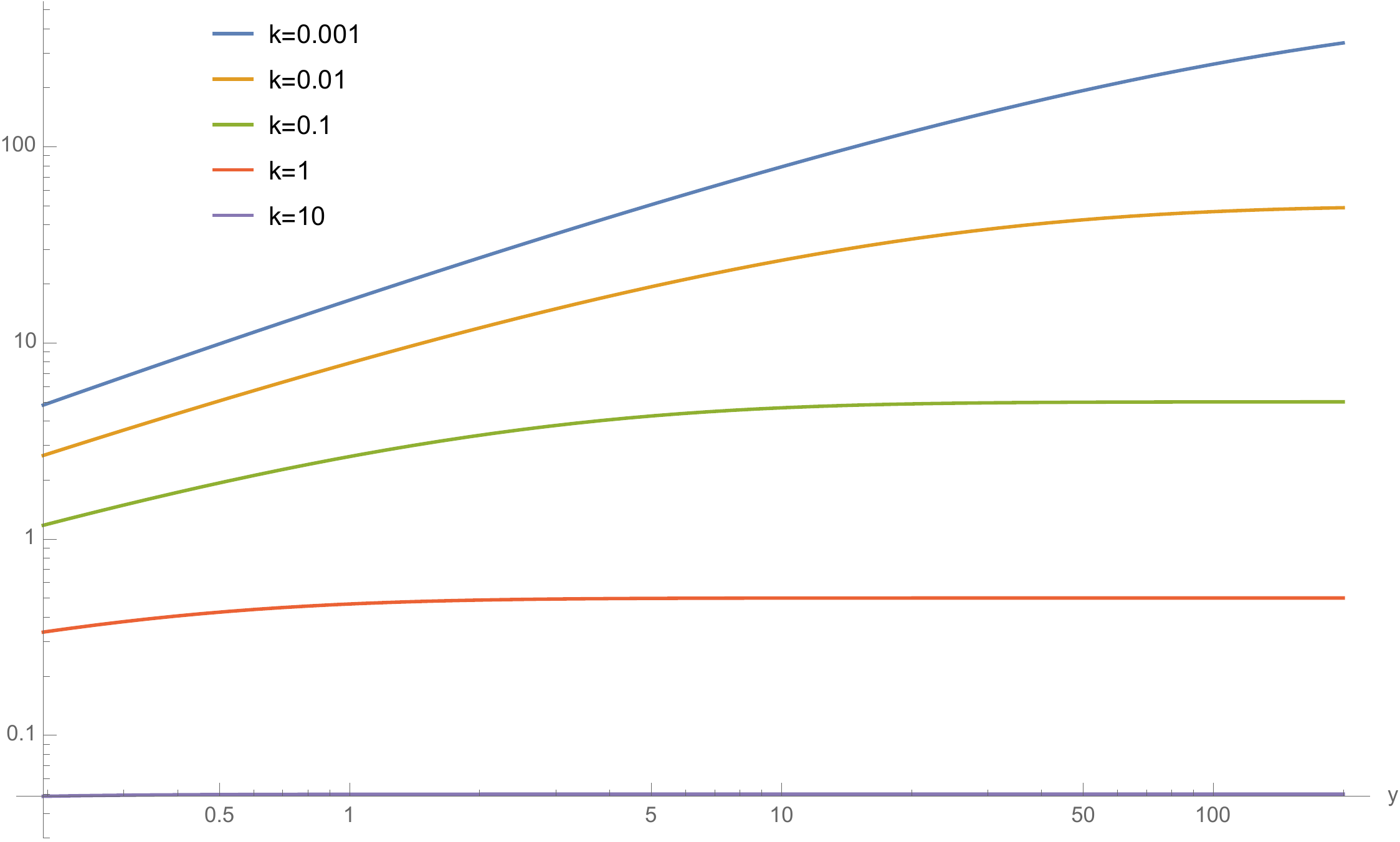}\includegraphics[width=7.5 cm]{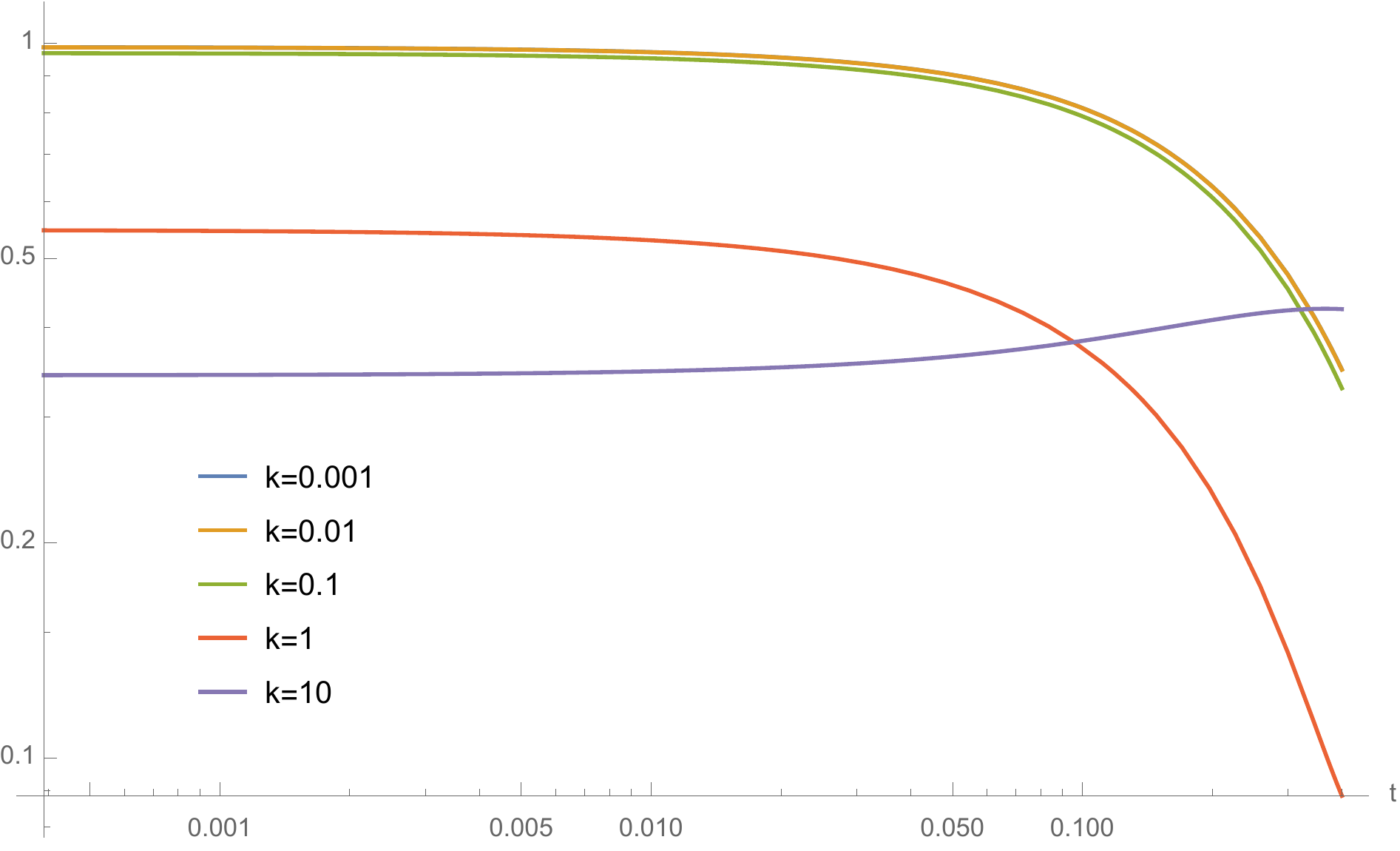}
\caption{{\emph{Left panel}}: Squared norm of the positive-frequency solutions \eqref{app2eq1} selected by our asymptotic expansion of $h_k$ in the kinetically dominated epoch of a pure GR cosmology with fast-roll inflation, as a function of the displaced conformal time $y$. {\emph{Right panel}}: Squared norm of the positive-frequency solutions \eqref{app2eq2} selected by our asymptotic expansion of $h_k$ in the vicinity of the LQC bounce with kinetic domination, as a function of the proper time $t\in[0,0.4]$.}
\label{fig5} 
\end{figure}

We show in figure \ref{fig5} the squared amplitude $|\mu_k|^2$ of the positive-frequency solutions in both of these cosmological scenarios, plotted as a function of the displaced conformal time $y$ for GR (up to an arbitrarily large value) and of the proper time for LQC. For both sets of solutions, that define the corresponding vacuum states, we see that no time-dependent or mode-dependent oscillations appear in the evolution of the amplitude.
\\

\acknowledgments

This work was partially supported by Project. No. MICINN FIS2017-86497-C2-2-P and Project. No. MICINN PID2020-118159GB-C41 from Spain. B.E.N. acknowledges financial support from the Standard program of JSPS Postdoctoral Fellowships for Research in Japan.


\end{document}